\documentclass[journal,comsoc]{IEEEtran}
\usepackage[T1]{fontenc}

\usepackage{color}
\usepackage{setspace}
\usepackage{cite}

\ifCLASSINFOpdf
  \usepackage[pdftex]{graphicx}
\else
\fi
\usepackage{amsmath}
\usepackage[cmintegrals]{newtxmath}
\usepackage{algorithmic}

\usepackage{array}
\usepackage{url}

\begin{document}
%
\title{CA-Net: Comprehensive Attention Convolutional Neural Networks for Explainable Medical Image Segmentation}
%
%
%

\author{Ran~Gu, Guotai~Wang, Tao~Song, Rui~Huang, Michael~Aertsen, Jan~Deprest, S\'ebastien~Ourselin, Tom~Vercauteren, Shaoting~Zhang
\thanks{This work was supported by National Natural Science Foundation of China. No.81771921, and No.61901084. (Corresponding author: Guotai~Wang)}
\thanks{R. Gu, G. Wang, and S. Zhang are with the School of Mechanical and Electrical Engineering, University of Electronic Science and Technology of China, Chengdu, China. (e-mail: guotai.wang@uestc.edu.cn).}
\thanks{T. Song, and R. Huang are with the SenseTime Research, Shanghai, China.}
\thanks{M. Aertsen is with the Department of Radiology, University Hospitals Leuven, Leuven, Belgium.}
\thanks{J. Deprest is with the School of Biomedical Engineering and Imaging Sciences, King's College London, London, U.K., with the Department of Obstetrics and Gynaecology, University Hospitals Leuven, Leuven, Belgium, and with the Institute for Women’s Health, University College London, London, U.K.}
\thanks{S. Ourselin, and T. Vercauteren are with the School of Biomedical Engineering and Imaging Sciences, King's College London, London, U.K.}
}
\maketitle

\begin{abstract}
Accurate medical image segmentation is essential for diagnosis and treatment planning of diseases. Convolutional Neural Networks (CNNs) have achieved state-of-the-art performance for automatic medical image segmentation. However, they are still challenged by complicated conditions where the segmentation target has large variations of position, shape and scale, and existing CNNs have a poor explainability that limits their application to clinical decisions.
In this work, we make extensive use of multiple attentions in a CNN architecture and propose a comprehensive attention-based CNN (CA-Net) for more accurate and explainable medical image segmentation 
that is aware of the most important spatial positions, channels and scales at the same time. In particular, we first propose a joint spatial attention module to make the network focus more on the foreground region. Then, a novel channel attention module is proposed to adaptively recalibrate channel-wise feature responses and highlight the most relevant feature channels. Also, we propose a scale attention module implicitly emphasizing the most salient feature maps among multiple scales so that the CNN is adaptive to the size of an object. Extensive experiments on skin lesion segmentation from ISIC 2018 and multi-class segmentation of fetal MRI found that our proposed CA-Net significantly improved the average segmentation Dice score from $87.77\%$ to $92.08\%$ for skin lesion, $84.79\%$ to $87.08\%$ for the placenta and $93.20\%$ to $95.88\%$ for the fetal brain respectively compared with U-Net. It reduced the model size to around 15 times smaller with close or even better accuracy compared with state-of-the-art DeepLabv3+. In addition, it has a much higher explainability than existing networks by visualizing the attention weight maps. Our code is available at https://github.com/HiLab-git/CA-Net
\end{abstract}


%

\section{Introduction}
%
%
%
%
\IEEEPARstart{A}{utomatic} medical image segmentation is important for facilitating quantitative pathology assessment, treatment planning and monitoring disease progression~\cite{litjens2017survey}. However, this is a challenging task due to several reasons. First, medical images can be acquired with a wide range of protocols and usually have low contrast and inhomogeneous appearances, leading to over-segmentation and under-segmentation~\cite{wang2018interactive}. Second, some structures have large variation of scales and shapes such as skin lesion in dermoscopic images~\cite{codella2019skin}, making it hard to construct a prior shape model. In addition, some structures may have large variation of position and orientation in a large image context, such as the placenta and fetal brain in Magnetic Resonance Imaging (MRI)~\cite{wang2018deepigeos,salehi2018real,wang2018interactive}. To achieve good segmentation performance, it is highly desirable for automatic segmentation methods to be aware of the scale and position of the target.

With the development of deep Convolutional Neural Networks (CNNs), state-of-the-art performance has been achieved for many segmentation tasks~\cite{litjens2017survey}. Compared with traditional methods, CNNs have a higher representation ability and can learn the most useful features automatically from a large dataset. However, most existing CNNs are faced with the following problems: Firstly, by design of the convolutional layer, they use shared weights at different spatial positions, which may lead to a lack of spatial awareness and thus have reduced performance when dealing with structures with flexible shapes and positions, especially for small targets. Secondly, they usually use a very large number of feature channels, while these channels may be redundant. Many networks such as the U-Net~\cite{ronneberger2015u} use a concatenation of low-level and high-level features with different semantic information. They may have different importance for the segmentation task, and highlighting the relevant channels while suppressing some irrelevant channels would benefit the segmentation task~\cite{hu2018squeeze}. Thirdly, CNNs usually extract multi-scale features to deal with objects at different scales but lack the awareness of the most suitable scale for a specific image to be segmented~\cite{chen2017deeplab}. Last but not least, the decisions of most existing CNNs are hard to explain and employed in a black box manner due to their nested non-linear structure, which limits their application to clinical decisions.

To address these problems, attention mechanism is promising for improving CNNs' segmentation performance as it mimics the human behavior of focusing on the most relevant information in the feature maps while suppressing irrelevant parts. Generally, there are different types of attentions that can be exploited for CNNs, such as paying attention to the relevant spatial regions, feature channels and scales. As an example of spatial attention, the Attention Gate (AG)~\cite{Oktay2018a} generates soft region proposals implicitly and highlights useful salient features for the segmentation of abdominal organs. The Squeeze and Excitation (SE) block~\cite{hu2018squeeze} is one kind of channel attention and it recalibrates useful channel feature maps related to the target. Qin~\cite{qin2018autofocus} used an attention to deal with multiple parallel branches with different receptive fields for brain tumor segmentation, and the same idea was used in prostate segmentation from ultrasound images~\cite{Wang2018deep}. However, these works have only demonstrated the effectiveness of using a single or two attention mechanisms for segmentation that may limit the performance and explainability of the network. We assume that taking a more comprehensive use of attentions would boost the segmentation performance and make it easier to understand how the network works.

For artificial intelligence systems, the explainability is highly desirable when applied to medical diagnosis~\cite{samek2017explainable}. The explainability of CNNs has a potiential for verification of the prediction, where the reliance of the networks on the correct features must be guaranteed~\cite{samek2017explainable}. It can also help human understand the model's weaknesses and strengths in order to improve the performance and discover new knowledge distilled from a large dataset.
In the segmentation task, explainability helps developers interpret and understand how the decision is obtained, and accordingly modify the network in order to gain better accuracy. Some early works tried to understand CNNs' decisions by visualizing feature maps or convolution kernels in different layers~\cite{chen2019explaining}. Other methods such as Class Activation Map (CAM)~\cite{zhou2016learning} and Guided Back Propagation (GBP)~\cite{springenberg2014striving} are mainly proposed for explaining decisions of CNNs in classification tasks. However, explainability of CNNs in the context of medical image segmentation has rarely been investigated~\cite{schlemper2019attention,roy2018concurrent}. Schlemper et al.~\cite{schlemper2019attention} proposed attention gate that implicitly learn to suppress irrelevant region while highlighting salient features. Furthermore, Roy et al.~\cite{roy2018concurrent} introduced spatial and channel attention at the same time to boost meaningful features. In this work, we take advantages of spatial, channel and scale attentions to interpret and understand how the pixel-level predictions are obtained by our network. Visualizing the attention weights obtained by our network not only helps to understand which image region is activated for the segmentation result, but also sheds light on the scale and channel that contribute most to the prediction.

To the best of our knowledge, this is the first work on using comprehensive attentions to  improve the performance and explainability of CNNs for medical image segmentation. The contribution of this work is three-fold. First, we propose a novel Comprehensive Attention-based Network (i.e., CA-Net) in order to make a complete use of attentions to spatial positions, channels and scales. Second, to implement each of these attentions, we propose novel building blocks including a dual-pathway multi-scale spatial attention module,  a novel residual channel attention module and a scale attention module that adaptively selects features from the most suitable scales. Thirdly, we use the comprehensive attention to obtain good explainability of our network where the segmentation result can be attributed to the relevant spatial areas, feature channels and scales. Our proposed CA-Net was validated on two segmentation tasks: binary skin lesion segmentation from dermoscipic images and multi-class segmentation of fetal MRI (including the fetal brain and the placenta), where the objects vary largely in position, scale and shape. Extensive experiments show that CA-Net outperforms its counterparts that use no or only partial attentions. In addition, by visualizing the attention weight maps, we achieved a good explainability of how CA-Net works for the segmentation tasks.
\begin{figure*}
    \centering
    \vspace*{-0.2cm}
    \setlength{\abovecaptionskip}{-0cm}
    \setlength{\belowcaptionskip}{-0.6cm}
    \includegraphics[width=0.7\textwidth]{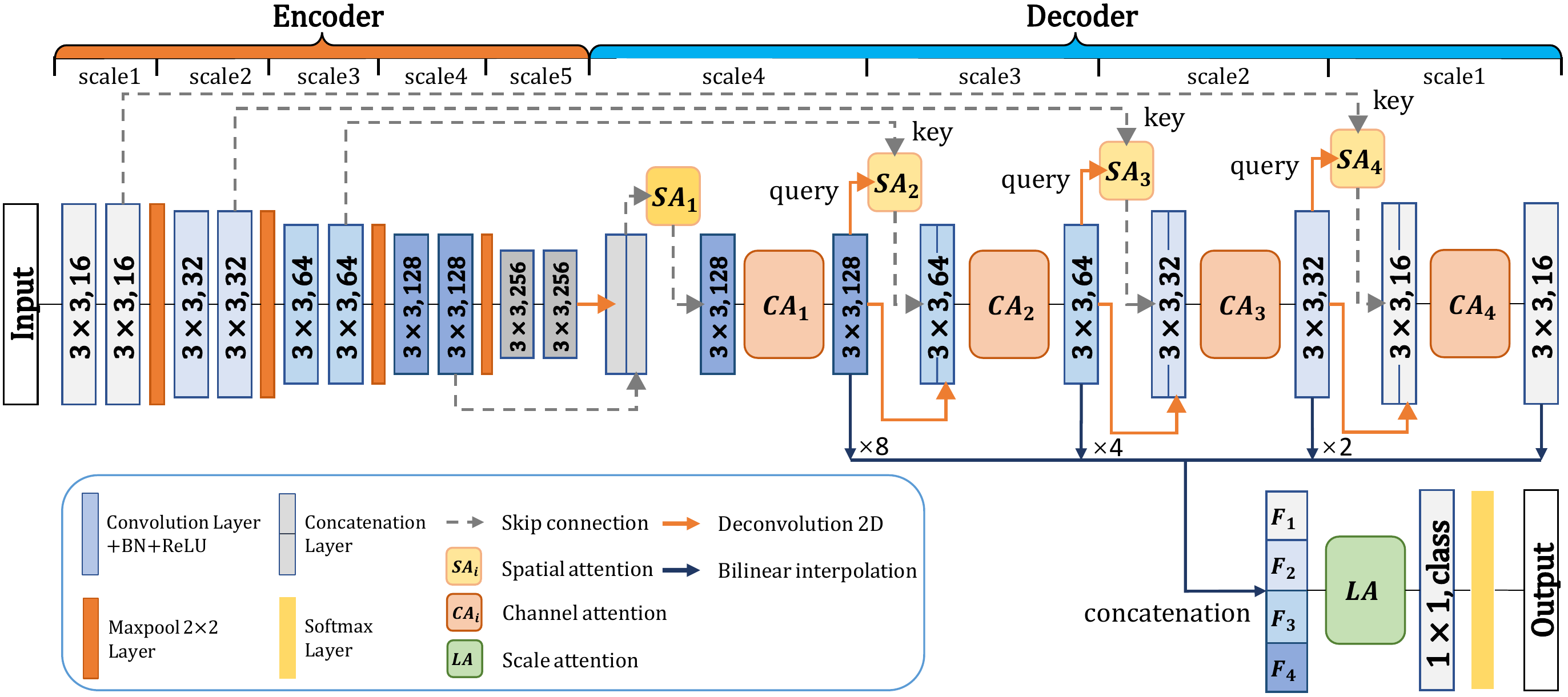}
    \caption{Our proposed comprehensive attention CNN (CA-Net). Blue rectangles with $3\times3$ or $1\times1$ and numbers (16, 32, 64, 128, and 256, or class) correspond to the convolution kernel size and the output channels. We use four spatial attentions ($SA_1$ to $SA_4$), four channel attentions ($CA_1$ to $CA_4$) and one scale attention ($LA$). $F_{1-4}$ means the resampled version of feature maps that are concatenated as input of the scale attention module.}
    \label{fig1:full_attention}
\end{figure*}
\section{Related Works}
\label{sec:related works}
\subsection{CNNs for Image Segmentation}
Fully Convolutional Network (FCN)~\cite{long2015fully} frameworks such as DeepLab~\cite{chen2017deeplab} are successful methods for natural semantic image segmentation. Subsequently, an encoder-decoder network SegNet~\cite{badrinarayanan2017segnet} was proposed to produce dense feature maps. DeepLabv3+~\cite{chen2018encoder} extended DeepLab by adding a decoder module and using depth-wise separable convolution for better performance and efficiency.

In medical image segmentation, FCNs also have been extensively exploited in a wide range of  tasks. 
U-Net~\cite{ronneberger2015u} is a widely used CNN for 2D biomedical image segmentation. The 3D U-Net~\cite{cciccek20163d} and V-Net~\cite{milletari2016v} with similar structures were proposed for 3D medical image segmentation. In~\cite{sarker2018slsdeep}, a dilated residual and pyramid pooling network was proposed for automated segmentation of melanoma. Some other CNNs with good performance for medical image segmentation include HighRes3DNet~\cite{li2017compactness}, DeepMedic~\cite{Kamnitsas2017a}, and H-DenseUNet~\cite{li2018tell}, etc. However, these methods only use position-invariant kernels for learning, without focusing on the features and positions that are more relevant to the segmentation object. Meanwhile, they have a poor explainability as they provide little mechanism for interpreting the decision-making process.
\subsection{Attention Mechanism}
In computer vision, there are attention mechanisms applied in different task scenarios~\cite{wang2017residual, fu2019dual, lu2017knowing}.
Spatial attention has been used for image classification~\cite{wang2017residual} and image caption~\cite{lu2017knowing}, etc. The learned attention vector highlights the salient spatial areas of the sequence conditioned on the current feature while suppressing the irrelevant counter-parts, making the prediction more contextualized. The SE block using a channel-wise attention was originally proposed for image classification and has recently been used for semantic segmentation~\cite{li2018tell,fu2019dual}. These ideas of attention mechanisms work by generating a context vector which assigns weights on the input sequence. In~\cite{chen2016attention}, an attention mechanism is proposed to lean to softly weight feature maps at multiple scales. However, this method feeds multiple resized input images to a shared deep network, which requires human expertise to choose the proper sizes and is not self-adaptive to the target scale.

Recently, to leverage attention mechanism for medical image segmentation, Oktay et al.~\cite{Oktay2018a} combined spatial attention with U-Net for abdominal pancreas segmentation from CT images. Roy et al.~\cite{roy2018concurrent} proposed concurrent spatial and channel wise 'Squeeze and Excitation' (scSE) frameworks for whole brain and abdominal multiple organs segmentation. Qin et al.~\cite{qin2018autofocus} and Wang et al.~\cite{Wang2018deep} got feature maps of different sizes from middle layers and recalibrate these feature maps by assigning an attention weight.
Despite the increasing number of works leveraging attention mechanisms for medical image segmentation, they seldom pay attention to feature maps at different scales. What's more, most of them focus on only one or two attention mechanisms, and to the best of our knowledge, the attention mechanisms have not been comprehensively incorporated to increase the accuracy and explainability of segmentation tasks.
\section{Methods}
\label{task1:methods}
\subsection{Comprehensive-Attention CNN}
The proposed CA-Net making use of comprehensive attentions is shown in Fig.~\ref{fig1:full_attention}, where we add specialized convolutional blocks to achieve comprehensive attention guidance with respect to the space, channel and scale of the feature maps simultaneously. Without loss of generality, we choose the powerful structure of the U-Net~\cite{ronneberger2015u} as the backbone. The U-Net backbone is an end-to-end-trainable network consisting of an encoder and a decoder with shortcut connection at each resolution level. The encoder is regarded as a feature extractor that obtains high-dimensional features across multiple scales sequentially, and the decoder utilizes these encoded features to recover the segmentation target.

Our CA-Net has four spatial attention modules ($SA_{1-4}$), four channel attention modules ($CA_{1-4}$) and one scale attention module ($LA$), as shown in Fig.~\ref{fig1:full_attention}. The spatial attention is utilized to strengthen the region of interest on the feature maps while suppressing the potential background or irrelevant parts. Hence, we propose a novel multi-scale spatial attention module that is a combination of non-local block~\cite{wang2018non} at the lowest resolution level ($SA_1$) and dual-pathway AG~\cite{Oktay2018a} at the other resolution levels ($SA_{2-4}$). We call it as the joint spatial attention ($Js-A$) that enhances inter-pixel relationship to make the network better focus on the segmentation target. Channel attention ($CA_{1-4}$) is used to calibrate the concatenation of low-level and high-level features in the network so that the more relevant channels are weighted by higher coefficients. Unlike the SE block that only uses average-pooling to gain channel attention weight, we additionally introduce max-pooled features to exploit more salient information for channel attention~\cite{Woo2018a}. 
Finally, we concatenate feature maps at multiple scales in the decoder and propose a scale attention module ($LA$) to highlight features at the most relevant scales for the segmentation target. These different attention modules are detailed in the following.
\subsubsection{Joint Spatial Attention Modules}
\label{}
The joint spatial attention is inspired by the non-local network~\cite{wang2018non} and AG~\cite{Oktay2018a}. We use four attention blocks ($SA_{1-4}$) in the network to learn attention maps at four different resolution levels, as shown in Fig.~\ref{fig1:full_attention}. 
\begin{figure*}[htb]
    \centering
    \vspace{-0.2cm}
    \setlength{\abovecaptionskip}{-0cm}
    \setlength{\belowcaptionskip}{-1cm}
    \includegraphics[width=0.96\textwidth]{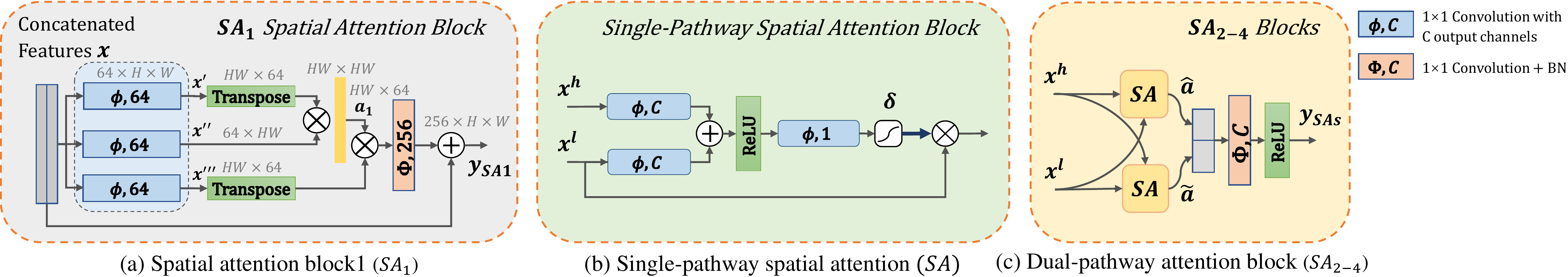}
    \caption{Details of our proposed joint spatial attention block. (a) $SA_1$ is a non-local block used at the lowest resolution level. (b) single-pathway spatial attention block (SA). (c) $SA_{2-4}$ are the dual-pathway attention blocks used in higher resolution levels. The \textit{query} feature $x^{h}$ is used to calibrate the low-level \textit{key} feature $x^{l}$. $\delta$ means the \textit{Sigmoid} function.}
    \label{fig2:spatial_atten}
\end{figure*}
First, for the spatial attention at the lowest resolution level ($SA_1$), we use a non-local block  that captures interactions between all pixels with a better awareness of the entire context. 
The detail of ($SA_{1}$) is shown in Fig.~\ref{fig2:spatial_atten}(a). 
Let $x$ represent the input feature map with a shape of $256\times H \times W$, where 256 is the input channel number, and $H$, $W$ represent the height and width, respectively. We first use three parallel $1 \times 1$ convolutional layers with an output channel number of 64 to reduce the dimension of $x$, obtaining three compressed feature maps $x'$, $x''$ and $x'''$, respectively, and they have the same shape of $64 \times H \times W$. The three feature maps can then be reshaped into 2D matrices with shape of $64 \times HW$. A spatial attention coefficient map is obtained as:
\begin{align}
    \alpha_1 = \sigma (x'^T\cdot x'')
\end{align}
where $T$ means matrix transpose operation.$\alpha_1\in (0, 1)^{HW \times HW}$ is a square matrix, and $\sigma$ is a row-wise \textit{Softmax} function so that the sum of each row equals to 1.0. $\alpha_1$ is used to represent the feature of each pixel as a weighted sum of features of all the pixels, to ensure the interaction among all the pixels. The calibrated feature map in the reduced dimension is:
\begin{align}
    \hat{x} = \alpha_1 \cdot x'''^{T}
\end{align}
~$\hat{x}$ is then reshaped to $64 \times H \times W$, and we use $\Phi^{256}$ that is a $1\times1$ convolution with batch normalization and output channel number of 256 to expand $\hat{x}$ to match the channel number of $x$. A residual connection is finally used to facilitate the information propagation during training, and the output of $SA_1$ is obtained as:
\begin{align}
    y_{SA1} = \Phi^{256}(\hat x) + x
\end{align}

Second, as the increased memory consumption limits applying the non-local block to feature maps with higher resolution, we extend AG to learn attention coefficients in $SA_{2-4}$. As a single AG may lead to a noisy spatial attention map, we propose a dual-pathway spatial attention that exploits two AGs in parallel to strengthen the attention to the region of interest as well as reducing noise in the attention map. Similarly to model ensemble, combining two AGs in parallel has a potential to improve the robustness of the segmentation. The details about a single pathway AG are shown in Fig.~\ref{fig2:spatial_atten}(b).
Let $x^{l}$ represent the low-level feature map at the scale $s$ in the encoder, and $x^{h}$ represent a high-level feature map up-sampled from the end of the decoder at scale $s+1$ with a lower spatial resolution, so that $x^{h}$ and $x^{l}$ have the same shape. 
In a single-pathway AG, the \textit{query} feature $x^{h}$ is used to calibrate the low-level \textit{key} feature $x^{l}$. As shown in Fig.~\ref{fig2:spatial_atten}(b), $x^{h}$ and $x^{l}$ are compressed by a $1\times 1$ convolution with an output channel number $C$ (e.g., 64) respectively, and the results are summed and followed by a \textit{ReLU} activation function. Feature map obtained by the \textit{ReLU} is then fed into another $1\times 1$ convlution with one output channel followed by a \textit{Sigmoid} function to obtain a pixel-wise attention coefficient $\alpha \in [0,1]^{H\times W}$. $x^{l}$ is then multiplied with $\alpha$ to be calibrated. In our dual-pathway AG, the  spatial attention maps in the two pathways are denoted as $\hat \alpha$ and $\tilde \alpha$ respectively. As shown in Fig.~\ref{fig2:spatial_atten}(c), the output of our dual-pathway AG for $SA_s (s=2,3,4)$ is obtained as: 
\begin{equation}
    y_{SAs} =\textit{ReLU}\big[\Phi^{C}\big( (x^l\cdot \hat \alpha) \copyright (x^l\cdot \tilde \alpha)\big)\big]
\end{equation}
where \copyright~means channel concatenation.
$\Phi^{C}$ denotes $1\times1$ convolution with $C$ output channels followed by batch normalization.
Here $C$ is 64, 32 and 16 for $SA_2$, $SA_3$ and $SA_4$, respectively.

\subsubsection{Channel Attention Modules}
\label{}
\begin{figure}[htb]
    \centering
    \vspace{-0.2cm}
    \setlength{\abovecaptionskip}{-0cm}
    \setlength{\belowcaptionskip}{-0.2cm}
    \includegraphics[width=0.48\textwidth]{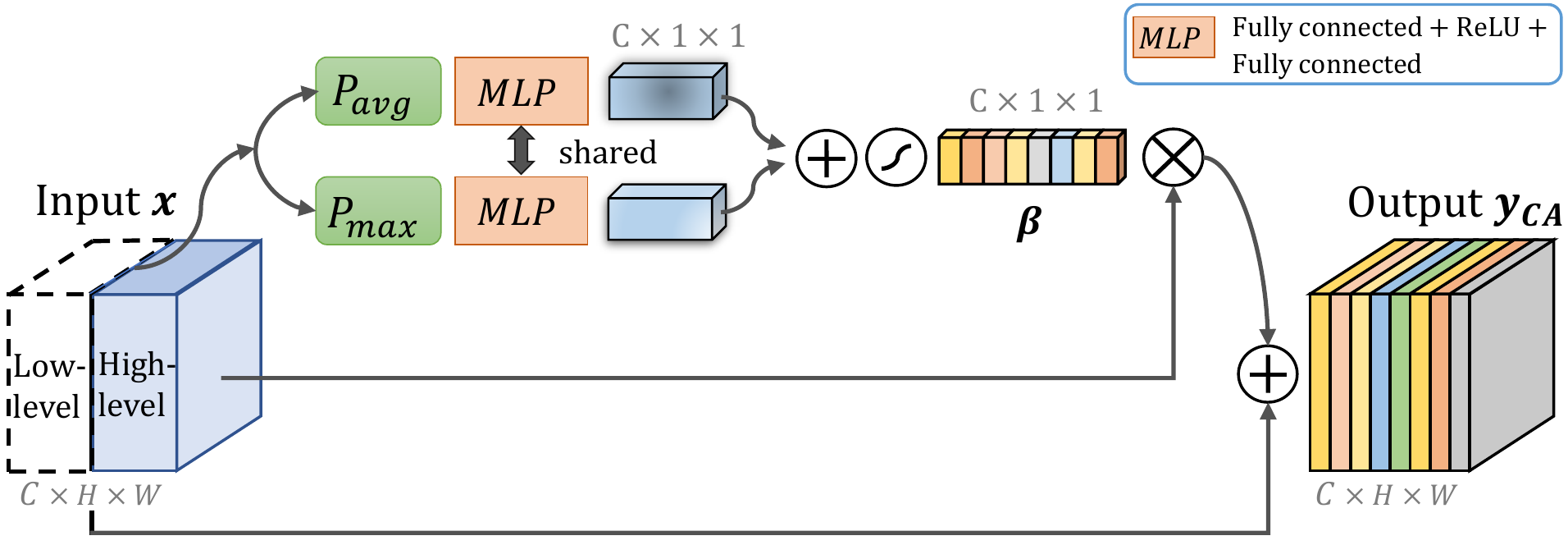}
    \caption{Structure of our proposed channel attention module with residual connection. Additional global max-pooled features are used in our module. $\beta$ means the channel attention coefficient.}
    \label{fig3:channel_atten}
\end{figure}

In our network, channel concatenation is used to combine the spatial attention-calibrated low-level features from the encoder and higher-level features from the decoder as shown in Fig.~\ref{fig1:full_attention}. Feature channels from the encoder contain mostly low-level information, and their counterparts from the decoder contain more semantic information. 
Therefore, they may have different importance for the segmentation task.~To better exploit the most useful feature channels, we introduce channel attention to automatically highlight the relevant feature channels while suppressing irrelevant channels. 
The details of proposed channel attention module ($CA_{1-4}$) is shown in Fig.~\ref{fig3:channel_atten}.

Unlike previous SE block that only utilized average-pooled information to excite feature channels~\cite{hu2018squeeze}, we use max-pooled features additionally to keep more information~\cite{Woo2018a}. Similarly, let $x$ represent the concatenated input feature map with $C$ channels, a global average pooling $P_{avg}$ and a global maximal pooling $P_{max}$ are first used to obtain the global information of each channel, and the outputs are represented as $P_{avg}(x)\in R^{C\times 1\times1}$ and $P_{max}(x)\in R^{C\times 1\times1}$, respectively. A multiple layer perception ($MLP$) $M^r$ is used to obtain the channel attention coefficient $\beta \in[0,1]^{C\times 1\times1}$, and $M^r$ is implemented by two fully connected layers, where the first one has an output channel number of $C/r$ followed by \textit{ReLU} and the second one has an output channel number of $C$. We set $r=2$ counting the trade-off of performance and computational cost~\cite{hu2018squeeze}. Note that a shared $M^r$ is used for $P_{avg}(x)$ and $P_{max}(x)$, and their results are summed and fed into a \textit{Sigmoid} to obtain $\beta$. The output of our channel attention module is obtained as:
\begin{equation}
    \setlength{\abovedisplayskip}{3pt}
    \setlength{\belowdisplayskip}{3pt}
    y_{CA}= x\cdot \beta + x
\end{equation}
where we use a residual connection to benefit the training. In our network, four channel attention modules $(CA_{1-4})$ are used (one for each concatenated feature), as shown in Fig.~\ref{fig1:full_attention}.
\subsubsection{Scale Attention Module}
\label{method:scale_atten}
The U-Net backbone obtains feature maps in different scales. To better deal with objects in different scales, it is reasonable to combine these features for the final prediction. However, for a given object, these feature maps at different scales may have different relevance to the object. It is desirable to automatically determine the scale-wise weight for each pixel, so that the network can be adaptive to corresponding scales of a given input. Therefore, we propose a scale attention module to learn image-specific weight for each scale automatically to calibrate the features at different scales, which is  used at the end of the network, as shown in Fig.~\ref{fig1:full_attention}. 
\begin{figure}[htb]
    \centering
    \vspace{-0.2cm}
    \setlength{\abovecaptionskip}{-0cm}
    \setlength{\belowcaptionskip}{-0.2cm}
    \includegraphics[width=0.49\textwidth]{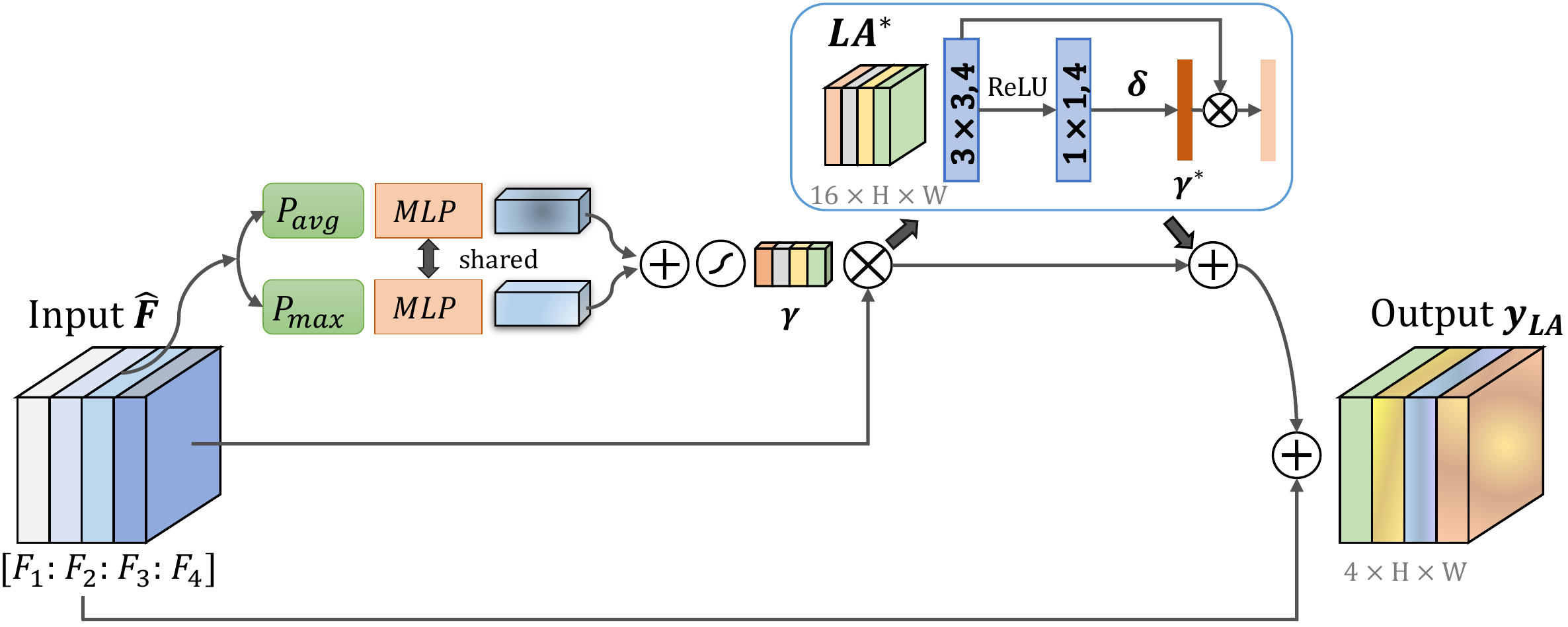}
    \caption{Structure of our proposed scale attention module with residual connection. Its input is the concatenation of interpolated feature maps at different scales obtained in the decoder. $\gamma$ means scale-wise attention coefficient. We additionally use a spatial attention block $LA^*$ to gain pixel-wise scale attention coefficient $\gamma^*$.}
    \label{fig4:scale_atten}
\end{figure}

Our proposed $LA$ block is illustrated in Fig.~\ref{fig4:scale_atten}. We first use bilinear interpolation to resample the feature maps $F_{s}$ at different scales ($s = 1, 2, 3, 4$) obtained by the decoder to the original image size. To reduce the computational cost, these feature map are compressed into four channels using $1\times1$ convolutions, and the compressed results from different scales are concatenated into a hybrid feature map $\hat{F}$. Similarly to our $CA$, we combine $P_{avg}$ $P_{max}$ with $MLP$ to obtain a coeffcieint for each channel (i.e., scale here), as shown in Fig.~\ref{fig4:scale_atten}. The scale coefficient attention vector is denoted as $\gamma \in[0,1]^{4\times1\times1}$. To distribute multi-scale soft attention weight on each pixel, we additionally use a spatial attention block $LA^*$ taking $\hat F \cdot \gamma $ as input to generate spatial-wise attention coefficient $\gamma^* \in[0,1]^{1\times H\times W}$, so that $\gamma \cdot\gamma^*$ represents a pixel-wise scale attention. $LA^*$ consists of one $3\times 3$ and one $1\times 1$ convolutional layers, where the first one has 4 output channels followed by \textit{ReLU}, and the second one has 4 output channels followed by \textit{Sigmoid}.
The final output of our $LA$ module is:
\begin{align}
    \setlength{\abovedisplayskip}{3pt}
    \setlength{\belowdisplayskip}{3pt}
    y_{LA} = \hat F \cdot \gamma \cdot\gamma^* +  \hat F \cdot \gamma  + \hat F
\end{align}
where the residual connections are again used to facilitate the training, as shown in Fig.~\ref{fig4:scale_atten}.  Using scale attention module enables the CNN to be aware of the most suitable scale (how big the object is).
\section{Experimental Results}
We validated our proposed framework with two applications: (i) Binary skin lesion segmentation from dermoscopic images. (ii) Multi-class segmentation of fetal MRI, including the fetal brain and the placenta. For both applications, we implemented ablation studies to validate the effectiveness of our proposed CA-Net and compared it with state-of-the-art networks. Experimental results of these two tasks will be detailed in Section~\ref{task1:skin_lesion} and Section~\ref{task2:fetal_mri}, respectively.
\begin{figure*}[htb]
    \centering
    \vspace{-0.4cm}
    \setlength{\abovecaptionskip}{-0cm}
    \setlength{\belowcaptionskip}{-0.4cm}
    \includegraphics[width=0.96\textwidth]{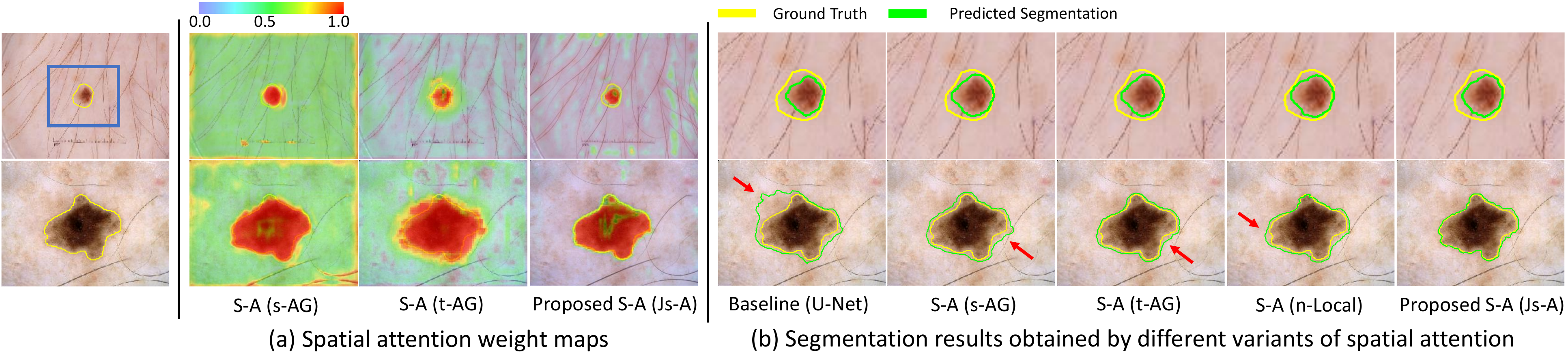}
    \caption{Visual comparison between different spatial attention methods for skin lesion segmentation. (a) is the visualized attention weight maps of single-pathway, dual-pathway and our proposed spatial attention. (b) shows segmentation results, where red arrows highlight some mis-segmentations. For better viewing of the segmentation boundary of the small target lesion, the first row of (b) shows the zoomed-in version of the region in the blue rectangle in (a).}
    \label{fig5:spatial_atten_result_with_atten_skin}
\end{figure*}
\subsection{Implementation and Evaluation Methods}
All methods were implemented in Pytorch framework\footnote{https://pytorch.org/}$^,$\footnote{https://github.com/HiLab-git/CA-Net}. We used Adaptive Moment Estimation (Adam) for training with initial learning rate $10^{-4}$, weight decay $10^{-8}$, batch size 16, and iteration 300 epochs. The learning rate is decayed by 0.5 every 256 epochs. The feature channel number in the first block of our CA-Net was set to 16 and doubled after each down-sampling. In $MLP$s of our $CA$ and $LA$ modules, the channel compression factor $r$ was 2 according to ~\cite{hu2018squeeze}. Training was implemented on one NVIDIA Geforce GTX 1080 Ti GPU.
We used Soft Dice loss function for the training of each network and used the best performing model on the validation set among all the epochs for testing. We used 5-fold cross-validation for final evaluation.
\begin{spacing}{1.0}
Quantitative evaluation of segmentation accuracy was based on: (i) The Dice score between a segmentation and the ground truth, which is defined as:
\begin{equation}
    Dice=\frac{2|\mathcal{R}_{a}\cap\mathcal{R}_{b}|}{|\mathcal{R}_{a}|+|\mathcal{R}_{b}|}
\end{equation}
where $\mathcal{R}_{a}$ and $\mathcal{R}_{b}$ denote the region segmented by algorithm and the ground truth, respectively. (ii) Average symmetric surface distance (ASSD). Supposing $S_{a}$ and $S_{b}$ represent the set of boundary points of the automatic segmentation and the ground truth respectively, the ASSD is defined as:
\end{spacing}
\begin{align}
    \mathit{ASSD}= \frac{1}{| S_{a}|+|S_{b}|} \times\bigg(\sum_{a\in S_{a}}d(a, S_{b})+\sum_{b\in S_{b}}d(b, S_{a})\bigg)
\end{align}
where $d(v, S_a)=\min_{w\in S_{a}}(\parallel v-w\parallel)$ denotes the minimum Euclidean distance from point $v$ to all the points of $S_a$.
\subsection{Lesion Segmentation from Dermoscopic Images}
\label{task1:skin_lesion}
With the emergence of automatic analysis algorithms, it becomes possible that accurate automatic skin lesion boundary segmentation helps dermatologists for early diagnosis and screening of skin diseases quickly. 
The main challenge for this task is that the skin lesion areas have various scales, shapes and colors, which requires automatic segmentation methods to be robust against shape and scale variations of the lesion~\cite{bi2017dermoscopic}.
\subsubsection{Dataset}
For skin lesion segmentation, we used the public available training set of ISIC 2018\footnote{https://challenge2018.isic-archive.com/} with 2594 images and their ground truth. We randomly split the dataset into 1816, 260 and 518 for training, validation and testing respectively. The original size of the skin lesion segmentation dataset ranged from $720\times540$ to $6708\times4439$, and we resized each image to $256\times342$ and normalized it by the mean value and standard deviation. During training, random cropping with a size of $224\times300$, horizontal and vertical flipping, and random rotation with a angle in $(-\pi/6, \pi/6)$ were used for data augmentation.

\subsubsection{Comparison of Spatial Attention Methods}
\label{spatial attention skin}
We first investigated the effectiveness of our spatial attention modules without using the channel attention and scale attention modules. We compared different variants of our proposed multi-level spatial attention: 1) Using standard single-pathway AG~\cite{Oktay2018a} at the position of $SA_{1-4}$, which is refereed to as s-AG; 2) Using the dual-pathway AG at the position of $SA_{1-4}$, which is refereed to as t-AG; 3) Using the non-local block of $SA_{1}$ only, which is refereed to as n-Local~\cite{wang2018non}. Our proposed joint attention method using non-local block in $SA_1$ and dual-pathway AG in $SA_{2-4}$ is denoted as Js-A. For the baseline U-Net, the skip connection was implemented by a simple concatenation of the corresponding features in the encoder and the decoder~\cite{ronneberger2015u}. For other compared variants that do not use $SA_{2-4}$, their skip connections were implemented as the same as that of U-Net.
Table~\ref{tab1:spatial_atten_table_skin} shows a quantitative comparison between these methods. It can be observed that all the variants using spatial attention lead to higher segmentation accuracy than the baseline. Also, we observe that dual-pathway spatial AG is more effective than single-pathway AG, and our joint spatial attention block outperforms the others. Compared with the standard AG~\cite{Oktay2018a}, our proposed spatial attention improved the average Dice from 88.46\% to 90.83\%.
\begin{table}[htb]
    \small
    \vspace{-0.2cm}
    \setlength{\abovecaptionskip}{0.2cm}
    \setlength{\belowcaptionskip}{-0.8cm}
    \centering
    \caption{Quantitative evaluation of different spatial attention methods for skin lesion segmentation. (s-AG) means single-pathway AG, (t-AG) means dual-pathway AG, (n-Local) means non-local networks. Js-A is our proposed multi-scale spatial attention that combines non-local block and dual-pathway AG.}
    \label{tab1:spatial_atten_table_skin}
    \scalebox{0.98}{\begin{tabular}{l|c|c|c}
        \hline
        Network & Para & Dice(\%) & ASSD(pix)\\
        \hline
        Baseline(U-Net~\cite{ronneberger2015u}) & 1.9M & 87.77$\pm$3.51 & 1.23$\pm$1.07 \\
        S-A(s-AG)~\cite{Oktay2018a} & 2.1M & 88.46$\pm$3.37 & 1.18$\pm$1.24 \\
        S-A(t-AG) & 2.4M & 89.18$\pm$3.29 & 0.90$\pm$0.50 \\
        S-A(n-Local) & 1.9M & 90.15$\pm$3.21 & \textbf{0.65$\pm$0.72} \\
        \textbf{Proposed S-A(Js-A)} & 2.0M & \textbf{90.83$\pm$3.31} & 0.81$\pm$1.06 \\ \hline
    \end{tabular}}
\end{table}
\begin{figure*}[htb]
    \centering
    \vspace{-0.4cm}
    \setlength{\abovecaptionskip}{-0cm}
    \setlength{\belowcaptionskip}{-0.4cm}
    \includegraphics[width=0.74\textwidth]{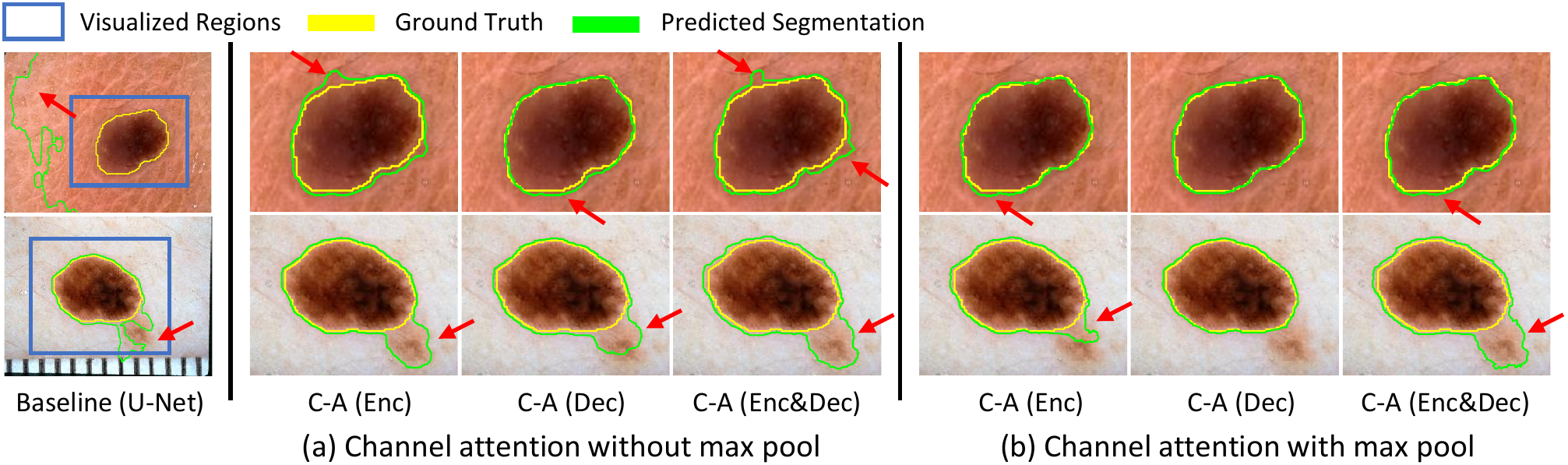}
    \caption{Visual comparison of our proposed channel attention method with different variants. Our proposed attention block is CA (Dec) where the channel attention module uses an additional max pooling and is plugged in the decoder. The red arrows highlight some mis-segmentations.}
    \label{fig6:channel_atten_result_skin}
\end{figure*}

Fig.~\ref{fig5:spatial_atten_result_with_atten_skin}(a) visualizes the spatial attention weight maps obtained by s-AG, t-AG and our Js-A. It can be observed that single-pathway AG pays attention to almost every pixel, which means it is dispersive. The dual-pathway AG is better than than the single-pathway AG but still not self-adaptive enough. In comparison, our proposed Js-A pays a more close attention to the target than the above methods.

Fig.~\ref{fig5:spatial_atten_result_with_atten_skin}(b) presents some examples of qualitative segmentation results obtained by the compared methods. It can be seen that introducing spatial attention block in neural network largely improves the segmentation accuracy. Furthermore, the proposed Js-A gets better result than the other spatial attention methods in both cases. In the second case where the lesion has a complex shape and blurry boundary, our proposed Js-A keeps a better result.

We observed that there may exist skew between original annotation and our cognition in ISIC 2018, as shown in Fig.~\ref{fig5:spatial_atten_result_with_atten_skin}. This is mainly because that the image contrast is often low along the true boundary, and the exact lesion boundary requires some expertise to delineate. The ISIC 2018 dataset was annotated by experienced dermatologists, and some annotations may be different from what a non-expert thinks.
\subsubsection{Comparison of Channel Attention Methods}
\label{task2:channel_atten_skin}
In this comparison, we only introduced channel attention modules to verify the effectiveness of our proposed method. We first investigate the effect of position in the network the channel attention module plugged in: 1) the encoder, 2) the decoder, 3) both the encoder and decoder. These three variants are referred to as C-A (Enc), C-A (Dec) and C-A (Enc\&Dec) respectively. We also compared the impact of using and not using max pooling for the channel attention module.
\begin{table}[htb]
    \small
    \vspace{-0.4cm}
    \setlength{\abovecaptionskip}{0.2cm}
    \setlength{\belowcaptionskip}{-0.4cm}
    \centering
    \caption{Quantitative comparison of different channel attention methods for skin lesion segmentation. Enc, Dec and Enc$\&$Dec means channel attention blocks are plugged in the encoder, the decoder and both encoder and decoder, respectively.}
    \label{tab2:channel_atten_table_skin}
    \scalebox{0.96}{\begin{tabular}{l|c|c|c|c}
        \hline
       Network & P$_{max}$ & Para & Dice($\%$) & ASSD(pix) \\ \hline
        Baseline & - & 1.9M  & 87.77$\pm$3.51 & 1.23$\pm$1.07 \\
        C-A(Enc) & $\times$ & 2.7M & 91.06$\pm$3.17 & 0.73$\pm$0.56 \\
        C-A(Enc) & $\surd$ & 2.7M & 91.36$\pm$3.09 & 0.74$\pm$0.55 \\
        C-A(Dec) & $\times$ & 2.7M & 91.56$\pm$3.17 & 0.64$\pm$0.45 \\
        \textbf{C-A(Dec)} & $\surd$ & 2.7M & \textbf{91.68$\pm$2.97} & 0.65$\pm$0.54 \\
        C-A(Enc$\&$Dec) & $\times$ & 3.4M & 90.85$\pm$3.42 & 0.92$\pm$1.40 \\
        C-A(Enc$\&$Dec) & $\surd$ & 3.4M & 91.63$\pm$3.50 & \textbf{0.58$\pm$0.41} \\ \hline
    \end{tabular}}
\end{table}

Table~\ref{tab2:channel_atten_table_skin} shows the quantitative comparison of these variants, which demonstrates that channel attention blocks indeed improve the segmentation performance.
Moreover, channel attention block with additional max-pooled information generally performs better than those using average pooling only. Additionally, we find that channel attention block plugged in the decoder performs better than plugged into the encoder or both the encoder and decoder. The C-A (Dec) achieved an average Dice score of 91.68\%, which outperforms the others. 

Fig.~\ref{fig6:channel_atten_result_skin} shows the visual comparison of our proposed channel attention and its variants. The baseline U-Net has a poor performance when the background has a complex texture, and the channel attention methods improve the accuracy for these cases. Clearly, our proposed channel attention module C-A (Dec) obtains higher accuracy than the others.

\subsubsection{Comparison of Scale Attention Methods}
\label{task: scale_attention_skin}
In this comparison, we only introduced scale attention methods to verify the effectiveness of our proposed scale attention. Let L-A (1-K) denote the scale attention applied to the concatenation of feature maps from scale 1 to K as shown in Fig.~\ref{fig1:full_attention}. To investigate the effect of number of feature map scales on the segmentation, we compared our proposed method with K=2, 3, 4 and 5 respectively.
\begin{table}[htb]
    \small
    \vspace{-0.4cm}
    \setlength{\abovecaptionskip}{0.2cm}
    \setlength{\belowcaptionskip}{-0.4cm}
    \centering
    \caption{Quantitative evaluation of different scale-attention methods for skin lesion segmentation. L-A (1-K) represents the features from scale 1 to K were concatenated for scale attention.}
    \label{tab3:scale_atten_table_skin}
    \scalebox{0.96}{\begin{tabular}{l|c|c|c}
        \hline
        Network & Para & Dice(\%) & ASSD(pix) \\ \hline
        Baseline & 1.9M & 87.77$\pm$3.51 & 1.23$\pm$1.07 \\
        L-A(1-2) & 2.0M & 91.21$\pm$3.33 & 1.00$\pm$1.36 \\
        L-A(1-3) & 2.0M & 91.53$\pm$2.52 & 0.70$\pm$0.61 \\
        \textbf{L-A(1-4)} & 2.0M & \textbf{91.58$\pm$2.48} & \textbf{0.66$\pm$0.47} \\
        L-A(1-5) & 2.0M & 89.67$\pm$3.40 & 0.82$\pm$0.50 \\ \hline
    \end{tabular}}
\end{table}

\begin{figure}[htb]
    \centering
    \vspace{-0.2cm}
    \setlength{\abovecaptionskip}{-0cm}
    \setlength{\belowcaptionskip}{-0.4cm}
    \includegraphics[width=0.49\textwidth]{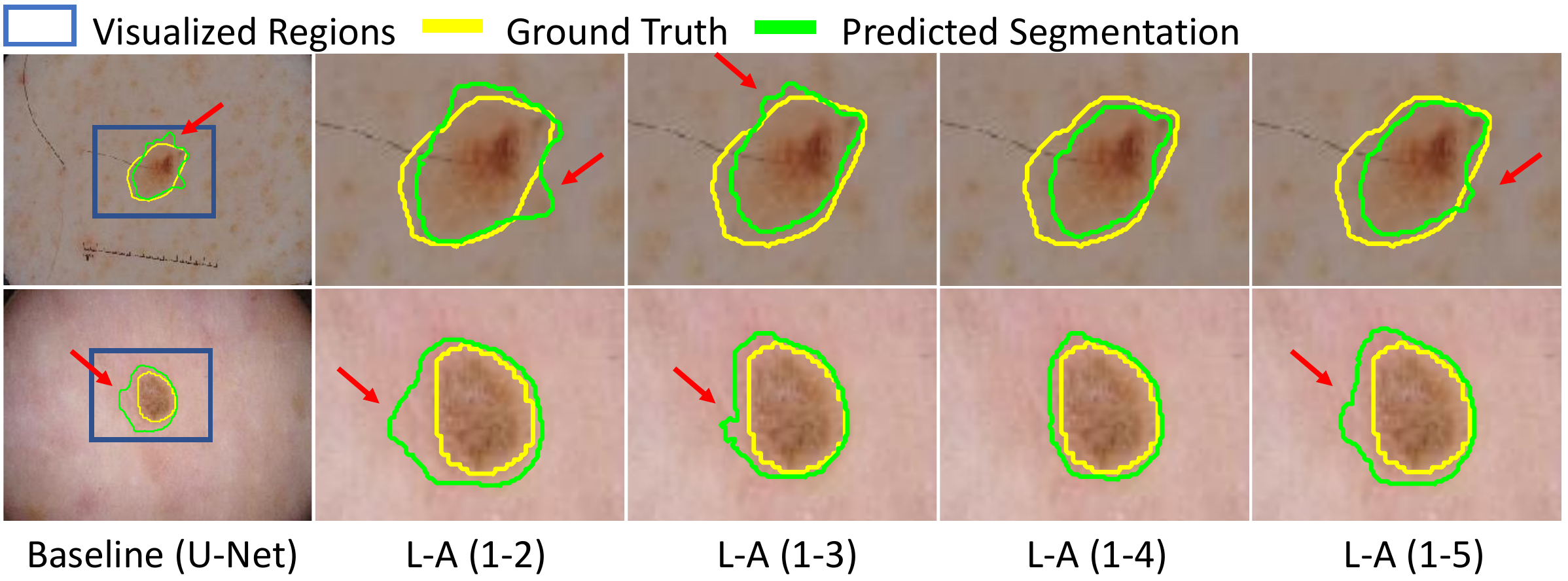}
    \caption{Visual comparison of segmentation obtained by scale attention applied to concatenation of features from different scales.}
    \label{fig7:scale_atten_result_skin}
\end{figure}

Table~\ref{tab3:scale_atten_table_skin} shows the quantitative comparison results. We find that combining features for multiple scales outperforms the baseline. When we concatenated features from scale 1 to 4, the Dice score and ASSD can get the best values of 91.58\% and 0.66 pixels respectively. However, when we combined features from all the 5 scales, the segmentation accuracy is decreased. This suggests that the feature maps at the lowest resolution level is not suitable for predicting pixel-wise label in details. As a result, we only fused the features from scale 1 to 4, as shown in Fig.~\ref{fig1:full_attention} in the following experiments. Fig.~\ref{fig7:scale_atten_result_skin} shows a visualization of skin lesion segmentation based on different scale attention variants. 
\begin{figure}[htb]
    \centering
    \vspace{-0.2cm}
    \setlength{\abovecaptionskip}{-0cm}
    \setlength{\belowcaptionskip}{-0.8cm}
    \includegraphics[width=0.49\textwidth]{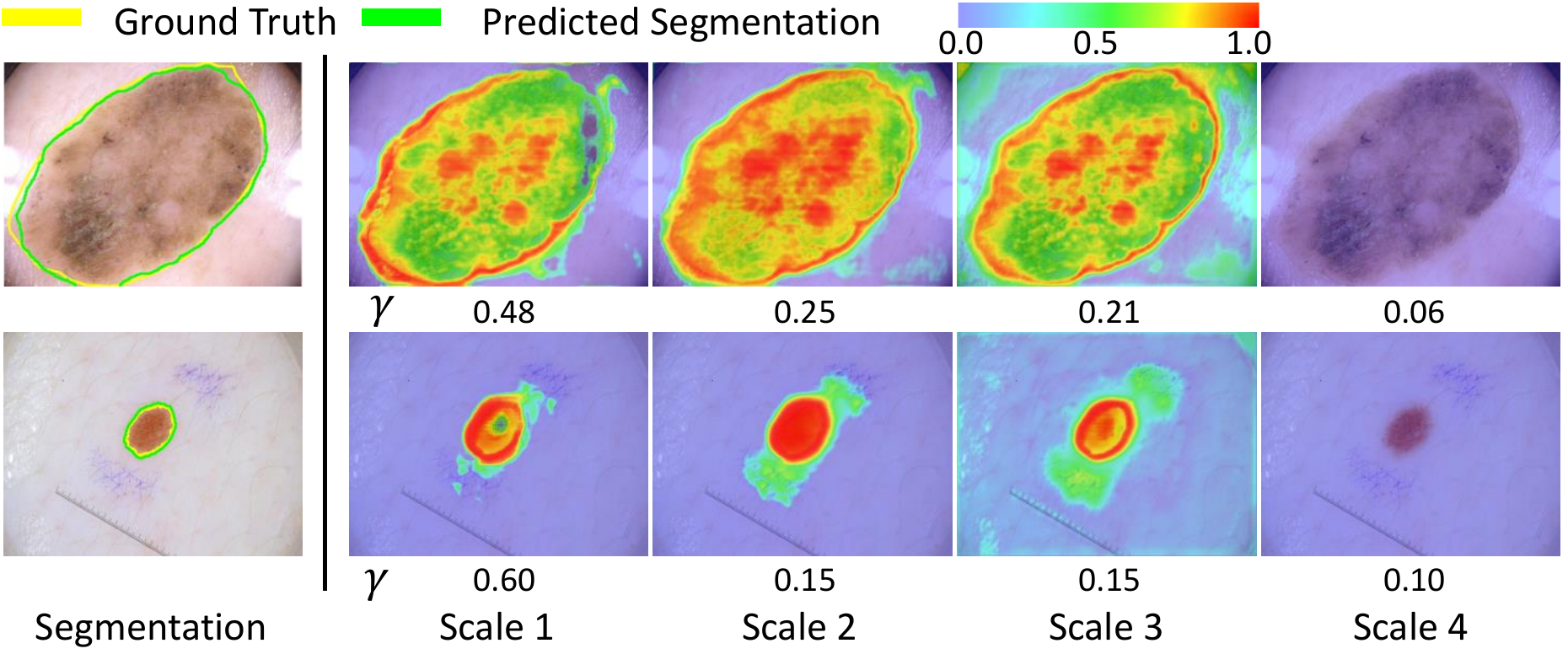}
    \caption{Visualization of scale attention on dermoscopic images. warmer color represents higher attention coefficient values. $\gamma$ means the global scale-wise attention coefficient.}
    \label{fig6_2:scale_atten_weight}
\end{figure}

Fig.~\ref{fig6_2:scale_atten_weight} presents the visualization of pixel-wise scale attention coefficient $\gamma\cdot \gamma^*$ at different scales, where the number under each picture denotes the scale-wise attention coefficient $\gamma$. This helps to better understand the importance of feature at different scales. The two cases show a large and a small lesion respectively. It can be observed that the large lesion has higher global attention coefficients $\gamma$ in scale 2 and 3 than the small lesion, and $\gamma$ in scale 1 has a higher value in the small lesion than the large lesion. The pixel-wise scale attention maps also show that the strongest attention is paid to scale 2 in the first row, and scale 1 in the second row. This demonstrates that the network automatically leans to focus on the corresponding scales for segmentation of lesions at different sizes.
\subsubsection{Comparison of Partial and Comprehensive Attention}
\label{partial attention skin}
To investigate the effect of combining different attention mechanisms, we compared CA-Net with six variants of different combinations of the three basic spatial, channel and scale attentions. Here, SA means our proposed multi-scale joint spatial attention and CA represents our channel attention used only in the decoder of the backbone. 
\begin{table}[htb]
    \small
    \vspace{-0.4cm}
    \setlength{\abovecaptionskip}{0.2cm}
    \setlength{\belowcaptionskip}{-0.6cm}
    \centering
    \caption{Comparison between partial and comprehensive attention methods for skin lesion segmentation. SA, CA and LA represent our proposed spatial, channel and scale attention modules respectively.}
    \label{tab5:full_partial_table_skin}
    \scalebox{0.92}{\begin{tabular}{l|c|c|c}
        \hline
        Network & Para & Dice(\%) & ASSD(pix) \\ \hline
        Baseline & 1.9M & 87.77$\pm$3.51 & 1.23$\pm$1.07 \\
        SA & 2.0M & 90.83$\pm$3.31 & 0.81$\pm$1.06 \\
        LA & 2.0M & 91.58$\pm$2.48 & 0.66$\pm$0.47 \\
        CA & 2.7M & 91.68$\pm$2.97 & 0.65$\pm$0.54 \\
        SA+LA & 2.1M & 91.62$\pm$3.13 & 0.70$\pm$0.48 \\
        CA+LA & 2.7M & 91.75$\pm$2.87 & 0.67$\pm$0.48 \\
        SA+CA & 2.8M & 91.87$\pm$3.00 & 0.73$\pm$0.69 \\
        \textbf{CA-Net(Ours)} & 2.8M & \textbf{92.08$\pm$2.67} & \textbf{0.58$\pm$0.39} \\ \hline
    \end{tabular}}
\end{table}
\begin{figure*}[htb]
    \centering
    \vspace{-0.4cm}
    \setlength{\abovecaptionskip}{-0cm}
    \setlength{\belowcaptionskip}{-0.4cm}
    \includegraphics[width=0.7\textwidth]{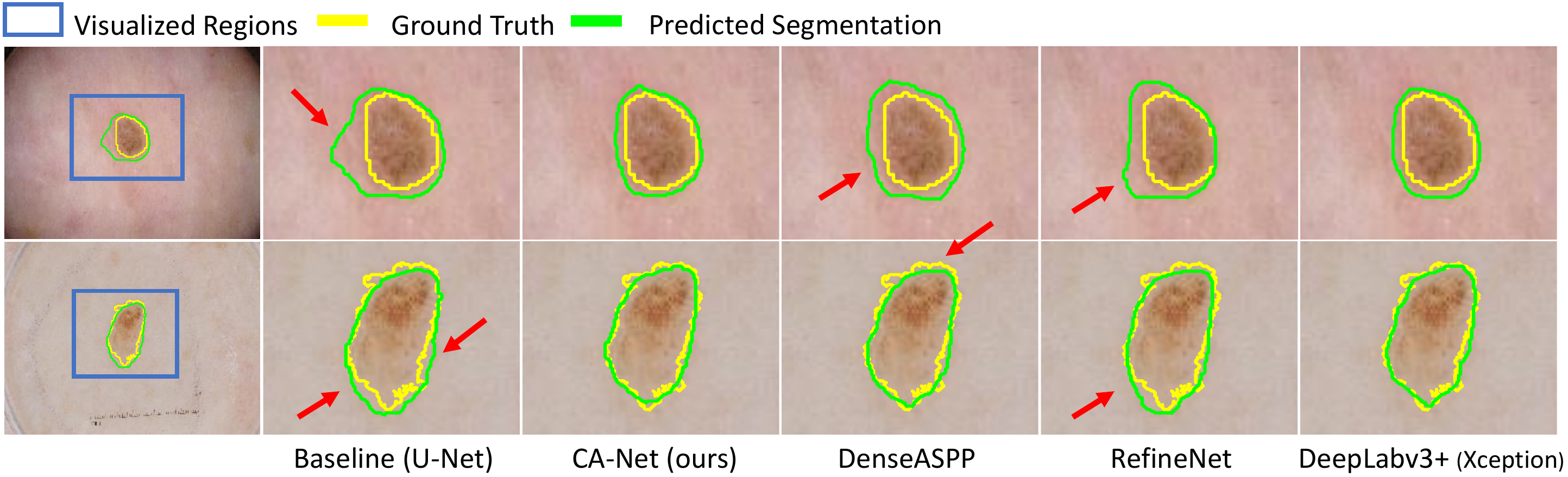}
    \caption{Visual comparison between CA-Net and state-of-the-art networks for skin lesion segmentation. Red arrows highlight some mis-segmentation. DeepLabv3+ has similar performance to ours, but our CA-Net has fewer parameters and better explainability.}
    \label{fig8:state_of_the_art_result_skin}
\end{figure*}

Table~\ref{tab5:full_partial_table_skin} presents the quantitative comparison of our CA-Net and partial attention methods for skin lesion segmentation. From Table~\ref{tab5:full_partial_table_skin}, we find that each of SA, CA and LA obtaines performance improvement compared with the baseline U-Net. Combining two of these attention methods outperforms the methods using a single attention. Furthermore, our proposed CA-Net outperforms all other variants both in Dice score and ASSD, and the corresponding values are 92.08\% and 0.58 pixels, respectively.
\subsubsection{Comparison with the State-of-the-Art Frameworks}
We compared our CA-Net with three state-of-the-art methods: 1) DenseASPP~\cite{yang2018denseaspp} that uses DenseNet-121~\cite{huang2017densely} as the backbone; 2) RefineNet~\cite{lin2017refinenet} that uses Resnet101~\cite{he2016deep} as the backbone; 3) Two variants of DeepLabv3+~\cite{chen2018encoder} that use Xception~\cite{chollet2017xception} and Dilated Residual Network (DRN)~\cite{yu2017dilated} as feature extractor, respectively. We retrained all these networks on ISIC 2018 and did not use their pre-trained models.
\begin{table}[htb]
    \small
    \vspace{-0.4cm}
    \setlength{\abovecaptionskip}{0.2cm}
    \setlength{\belowcaptionskip}{-0.4cm}
    \centering
    \caption{Comparison of the state-of-the-art methods and our proposed CA-Net for skin lesion segmentation. Inf-T means the inference time for a single image. E-able means the method is explainable.}
    \label{tab4:skin_fetal_table_skin}
    \scalebox{0.88}{\begin{tabular}{l|c|c|c|c}
        \hline
        Network & Para/Inf-T &\footnotesize{E-able}& Dice(\%) & ASSD(pix) \\ \hline
        \footnotesize{Baseline(U-Net~\cite{ronneberger2015u})}&1.9M/1.7ms&$\times$& 87.77$\pm$3.51 & 1.23$\pm$1.07 \\
        Attention U-Net~\cite{Oktay2018a} &2.1M/1.8ms&$\surd$& 88.46$\pm$3.37 & 1.18$\pm$1.24 \\
        \footnotesize{DenseASPP~\cite{yang2018denseaspp}}&8.3M/4.2ms&$\times$& 90.80$\pm$3.81 & 0.59$\pm$0.70 \\
        \footnotesize{DeepLabv3+(DRN)}& 40.7M/2.2ms&$\times$& 91.79$\pm$3.39 & 0.54$\pm$0.64 \\
        \footnotesize{RefineNet~\cite{lin2017refinenet}}&46.3M/3.4ms&$\times$& 91.55$\pm$2.11 & 0.64$\pm$0.77 \\
        \footnotesize{DeepLabv3+~\cite{chen2018encoder}}&54.7M/4.0ms&$\times$& 92.21$\pm$3.38 & 0.48$\pm$0.58 \\  \footnotesize{\textbf{CA-Net(Ours)}}&2.8M/2.1ms&$\surd$& 92.08$\pm$2.67 & 0.58$\pm$0.56 \\ \hline
    \end{tabular}}
\end{table}
\begin{table}
    \small
    \vspace{-0.4cm}
    \setlength{\abovecaptionskip}{0.2cm}
    \setlength{\belowcaptionskip}{-0.6cm}
    \centering
    \caption{Quantitative evaluation of different spatial attention methods for placenta and fetal brain segmentation. s-AG means single-pathway AG, t-AG means dual-pathway AG, n-Local means non-local networks. Js-A is our proposed multi-scale spatial attention that combines non-local block and dual-pathway AG.}
    \label{tab1:spatial_atten_table_mri}
    \scalebox{0.88}{\begin{tabular}{l|c|c|c|c}
        \hline
        \multicolumn{1}{l|}{Network} & \multicolumn{2}{c|}{Placenta} & \multicolumn{2}{c}{Fetal Brain} \\
        \cline{2-5}
        \multicolumn{1}{l|}{} & Dice(\%) & ASSD(pix) & Dice(\%) & ASSD(pix) \\ \hline
        Baseline & 84.79$\pm$8.45 & 0.77$\pm$0.95 & 93.20$\pm$5.96 & 0.38$\pm$0.92 \\
        s-AG~\cite{Oktay2018a} & 84.71$\pm$6.62 & 0.72$\pm$0.61 & 93.97$\pm$3.19 & 0.47$\pm$0.73 \\
        t-AG & 85.26$\pm$6.81 & 0.71$\pm$0.70 & 94.70$\pm$3.63 & \textbf{0.30$\pm$0.40} \\
        n-Local & 85.43$\pm$6.80 & 0.66$\pm$0.55 & 94.57$\pm$3.48 & 0.37$\pm$0.53 \\
        \textbf{Js-A} & \textbf{85.65$\pm$6.19} & \textbf{0.58$\pm$0.43} & \textbf{95.47$\pm$2.43} & 0.30$\pm$0.49 \\ \hline
    \end{tabular}}
\end{table}
\begin{figure*}[htb]
    \centering
    \vspace{-0.2cm}
    \setlength{\abovecaptionskip}{-0cm}
    \setlength{\belowcaptionskip}{-0.8cm}
    \includegraphics[width=0.96\textwidth]{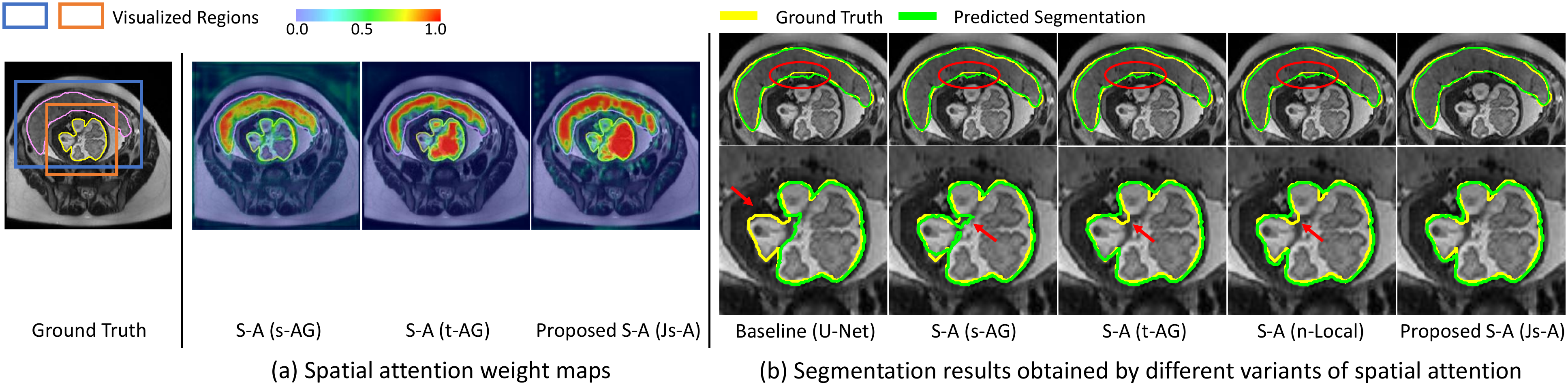}
    \caption{Visual comparison between different spatial attention methods for fetal MRI segmentation. (a) is the visualized attention weight maps of single-pathway, dual-pathway and proposed spatial attention. (b) shows segmentation results, where red arrows and circles highlight mis-segmentations.}
    \label{fig5:spatial_atten_result_with_atten_mri}
\end{figure*}

Quantitative comparison results of these methods are presented in Table~\ref{tab4:skin_fetal_table_skin}. It shows that all state-of-the-art methods have good performance in terms of Dice score and ASSD. 
Our CA-Net obtained a Dice score of 92.08\%, which is a considerable improvement compared with the U-Net whose Dice is 87.77\%. Though our CA-Net has a slightly lower performance than DeepLabv3+, the difference is not significant (p-value=0.46 $>$ 0.05), and our CA-Net has a model size that is 15 times smaller with better explainability. For skin lesion segmentation, the average inference time per image for our CA-Net was 2.1ms, compared with 4.0ms and 3.4ms by DeepLabv3+~\cite{chen2018encoder} and RefineNet~\cite{lin2017refinenet}, respectively. Fig.~\ref{fig8:state_of_the_art_result_skin} shows the visual comparison of different CNNs dealing with skin lesion segmentation task.

\subsection{Segmentation of Multiple Organs from Fetal MRI}
\label{task2:fetal_mri}
In this experiment, we demonstrate the effectiveness of our CA-Net in multi-organ segmentation, where we aim to jointly segment the placenta and the fetal brain from fetal MRI slices.
Fetal MRI has been increasingly used to study fetal development and pathology, as it provides a better soft tissue contrast than more widely used prenatal sonography~\cite{wang2018deepigeos}. Segmentation of some important organs such as the fetal brain and the placenta is important for fetal growth assessment and motion correction~\cite{torrents2019segmentation}. 
Clinical fetal MRI data are often acquired with a large slice thickness for good contrast-to-noise ratio.~Moreover, movement of the fetus can lead to inhomogeneous appearances between slices. Hence, 2D segmentation is considered more suitable than direct 3D segmentation from motion-corrupted MRI slices~\cite{wang2018interactive}.

\subsubsection{Dataset}
The dataset consists of 150 stacks with three views (axial, coronal, and sagittal) of T2-weighted fetal MRI scans of 36 pregnant women in the second trimester with Single-shot Fast-Spin echo (SSFSE) with pixel size 0.74 to 1.58 mm and inter-slice spacing 3 to 4 mm. The gestational age ranged from 22 to 29 weeks. 8 of the fetuses were diagnosed with spinal bifida and the others had no fetal pathologies. All the pregnant women were above 18 years old, and the use of data was approved by the Research Ethics Committee of the hospital. 

As the stacks contained an imbalanced number of slices covering the objects, we randomly selected 10 of these slices from each stack for the experiments. Then, we randomly split the slices at patient level and assigned 1050 for training, 150 for validation, and 300 for testing. The test set contains 110 axial slices, 80 coronal slices, and 110 sagittal slices. Manual annotations of the fetal brain and placenta by an experienced radiologist were used as the ground truth. We trained a multi-class segmentation network for simultaneous segmentation of these two organs. Each slice was resized to $256\times256$. We randomly flipped in x and y axis and rotated with an angle in $(-\pi/6, \pi/6)$ for data augmentation. All the images were normalized by the mean value and standard deviation.

\subsubsection{Comparison of Spatial Attention Methods}
In parallel to section~\ref{spatial attention skin}, we compared our proposed Js-A with: (1) the single-pathway AG (s-AG) only, (2) the dual-pathway AG (t-AG) only, (3) the non-local block (n-local) only.

Table~\ref{tab1:spatial_atten_table_mri} presents quantitative comparison results between these methods. From Table~\ref{tab1:spatial_atten_table_mri}, we observe that all the variants of spatial attention modules led to better Dice and ASSD scores. It can be observed that dual-pathway AG performs more robustly than the single-pathway AG, and Js-A module can get the highest scores, with Dice of 95.47\% and ASSD of 0.30 pixels, respectively. Furthermore, in placenta segmentation which has fuzzy tissue boundary, our model still maintains encouraging segmentation performance, with Dice score of 85.65\%, ASSD of 0.58 pixels, respectively.

Fig.~\ref{fig5:spatial_atten_result_with_atten_mri} presents a visual comparison of segmentation results obtained by these methods as well as their attention weight maps. From Fig.~\ref{fig5:spatial_atten_result_with_atten_mri}(b), we find that spatial attention has reliable performance when dealing with complex object shapes, as highlighted by red arrows. Meanwhile, with visualizing their spatial attention weight maps as shown in Fig.~\ref{fig5:spatial_atten_result_with_atten_mri}(a), our proposed Js-A has a greater ability to focus on the target areas compared with the other methods as it distributes a higher and closer weight on the target of our interest. 
\subsubsection{Comparison of Channel Attention Methods}
We compared the proposed channel attention method with the same variants as listed in section~\ref{task2:channel_atten_skin} for fetal MRI segmentation. The comparison results are presented in Table~\ref{tab2:channel_atten_table_mri}. It shows that channel attention plugged in decoder brings noticeably fewer parameters and still maintains similar or higher accuracy than the other variants. We also compared using and not using max-pooling in the channel attention block. From Table~\ref{tab2:channel_atten_table_mri}, we can find that adding extra max-pooled information indeed increases performance in terms of Dice and ASSD,  which proves the effectiveness of our proposed method.
\begin{table}[htb]
    \small
    \vspace{-0.4cm}
    \setlength{\abovecaptionskip}{0.2cm}
    \setlength{\belowcaptionskip}{-0.4cm}
    \centering
    \caption{Comparison experiment on channel attention-based networks for fetal MRI segmentation. Enc, Dec, and Enc\&Dec means the channel attention blocks are located in the encoder, decoder and both encoder and decoder, respectively.}
    \label{tab2:channel_atten_table_mri}
    \scalebox{0.725}{\begin{tabular}{l|c|c|c|c|c|c}
        \hline
        \multicolumn{1}{l|}{Network} & \multicolumn{1}{l|}{P$_{max}$} & \multicolumn{1}{l|}{Para} & \multicolumn{2}{c|}{Placenta} & \multicolumn{2}{c}{Fetal Brain} \\
        \cline{4-7}
        \multicolumn{1}{l|}{}  & \multicolumn{1}{l|}{} & \multicolumn{1}{l|}{} & Dice(\%) & ASSD(pix) & Dice(\%) & ASSD(pix) \\ \hline
        Baseline & - & 1.9M & 84.79$\pm$8.45 & 0.77$\pm$0.95 & 93.20$\pm$5.96 & 0.38$\pm$0.92 \\
        C-A(Enc) & $\times$ & 2.7M & 85.42$\pm$6.46 & 0.51$\pm$0.32 & 95.42$\pm$2.24 & 0.36$\pm$0.24 \\
        C-A(Enc) & $\surd$ & 2.7M & 86.12$\pm$7.00 & 0.50$\pm$0.44 & 95.60$\pm$3.30 & 0.31$\pm$0.36 \\
        C-A(Dec) & $\times$ & 2.7M & 86.17$\pm$5.70 & \textbf{0.44$\pm$0.29} & 95.61$\pm$3.69 & 0.33$\pm$0.42 \\
        \textbf{C-A(Dec)} & $\surd$ & 2.7M & \textbf{86.65$\pm$5.99} & 0.52$\pm$0.40 & \textbf{95.69$\pm$2.66} & 0.28$\pm$0.39 \\
        C-A(Enc$\&$Dec) & $\times$ & 3.4M & 85.83$\pm$7.02 & 0.52$\pm$0.40 & 95.60$\pm$2.29 & \textbf{0.26$\pm$0.46} \\
        C-A(Enc$\&$Dec) & $\surd$ & 3.4M & 86.26$\pm$6.68 & 0.51$\pm$0.43 & 94.39$\pm$4.14 & 0.47$\pm$0.64 \\ \hline
    \end{tabular}}
\end{table}
\subsubsection{Comparison of Scale Attention Methods}
In this comparison, we investigate the effect of concatenating different number of feature maps from scale 1 to K as described in section~\ref{task: scale_attention_skin}, and Table~\ref{tab3:scale_atten_table_mri} presents the quantitative results. Analogously, we observe that combining features from multiple scales outperforms the baseline. When we concatenate features from scale 1 to 4, we get the best results, and the corresponding Dice values for the placenta and the fetal brain are 86.21\% and 95.18\%, respectively. When the feature maps at the lowest resolution is additional used, i.e., L-A (1-5), the Dice scores are slightly reduced.

\begin{table}[htb]
    \small
    \vspace{-0.4cm}
    \setlength{\abovecaptionskip}{0.2cm}
    \setlength{\belowcaptionskip}{-0.4cm}
    \centering
    \caption{Comparison between different variants of scale attention-based networks. L-A (1-K) represents the features from scale 1 to K were concatenated for scale attention.}
    \label{tab3:scale_atten_table_mri}
    \scalebox{0.86}{\begin{tabular}{l|c|c|c|c}
        \hline
        \multicolumn{1}{l|}{Network} & \multicolumn{2}{c|}{Placenta} & \multicolumn{2}{c}{Fetal Brain} \\
        \cline{2-5}
        \multicolumn{1}{l|}{} & Dice(\%) & ASSD(pix) & Dice(\%) & ASSD(pix) \\ \hline
        Baseline & 84.79$\pm$8.45 & 0.77$\pm$0.95 & 93.20$\pm$5.96 & 0.38$\pm$0.92 \\
        L-A(1-2) & 86.17$\pm$6.02 & 0.59$\pm$0.41 & 94.19$\pm$3.29 & 0.27$\pm$0.36 \\
        L-A(1-3) & 86.17$\pm$6.53 & \textbf{0.50$\pm$0.37} & 94.61$\pm$3.13 & 0.54$\pm$0.63 \\
        \textbf{L-A(1-4)} & \textbf{86.21$\pm$5.96} & 0.52$\pm$0.58 & \textbf{95.18$\pm$3.22} & 0.27$\pm$0.59 \\
        L-A(1-5) & 86.09$\pm$6.10 & 0.61$\pm$0.47 & 95.05$\pm$2.51 & \textbf{0.24$\pm$0.47} \\ \hline
    \end{tabular}}
\end{table}
Fig.~\ref{fig7:scale_atten_result_mri} shows the visual comparison of our proposed scale attention and its variants. In the second row, the placenta has a complex shape with a long tail, and combining features from scale 1 to 4 obtained the best performance. Fig.~\ref{fig6_3:scale_atten_weight_mri} shows the visual comparison of scale attention weight maps on fetal MRI. From the visualized pixel-wise scale attention maps, we observed that the network pays much attention to scale 1 in the first row where the fetal brain is small, and to scale 2 in the second row where the fetal brain is larger. 
\begin{figure}[htb] 
    \vspace{-0.4cm}
    \setlength{\abovecaptionskip}{-0cm}
    \setlength{\belowcaptionskip}{-0.6cm}
    \centering
    \includegraphics[width=0.49\textwidth]{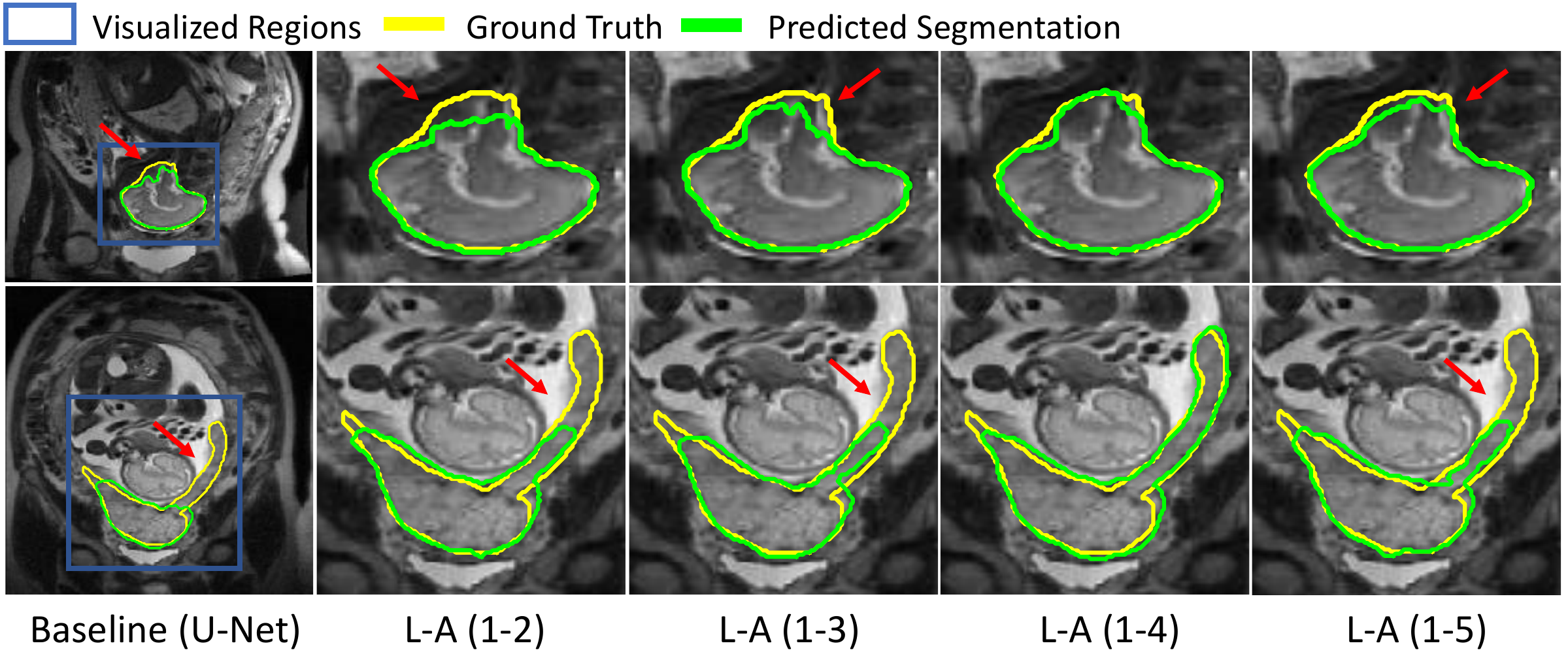}
    \caption{Visual comparison of proposed scale attention method applied to concatenation of features form different scales.}
    \label{fig7:scale_atten_result_mri}
\end{figure}
\begin{figure}[htb]
    \vspace{-0.2cm}
    \setlength{\abovecaptionskip}{-0cm}
    \setlength{\belowcaptionskip}{-1cm}
    \centering
    \includegraphics[width=0.49\textwidth]{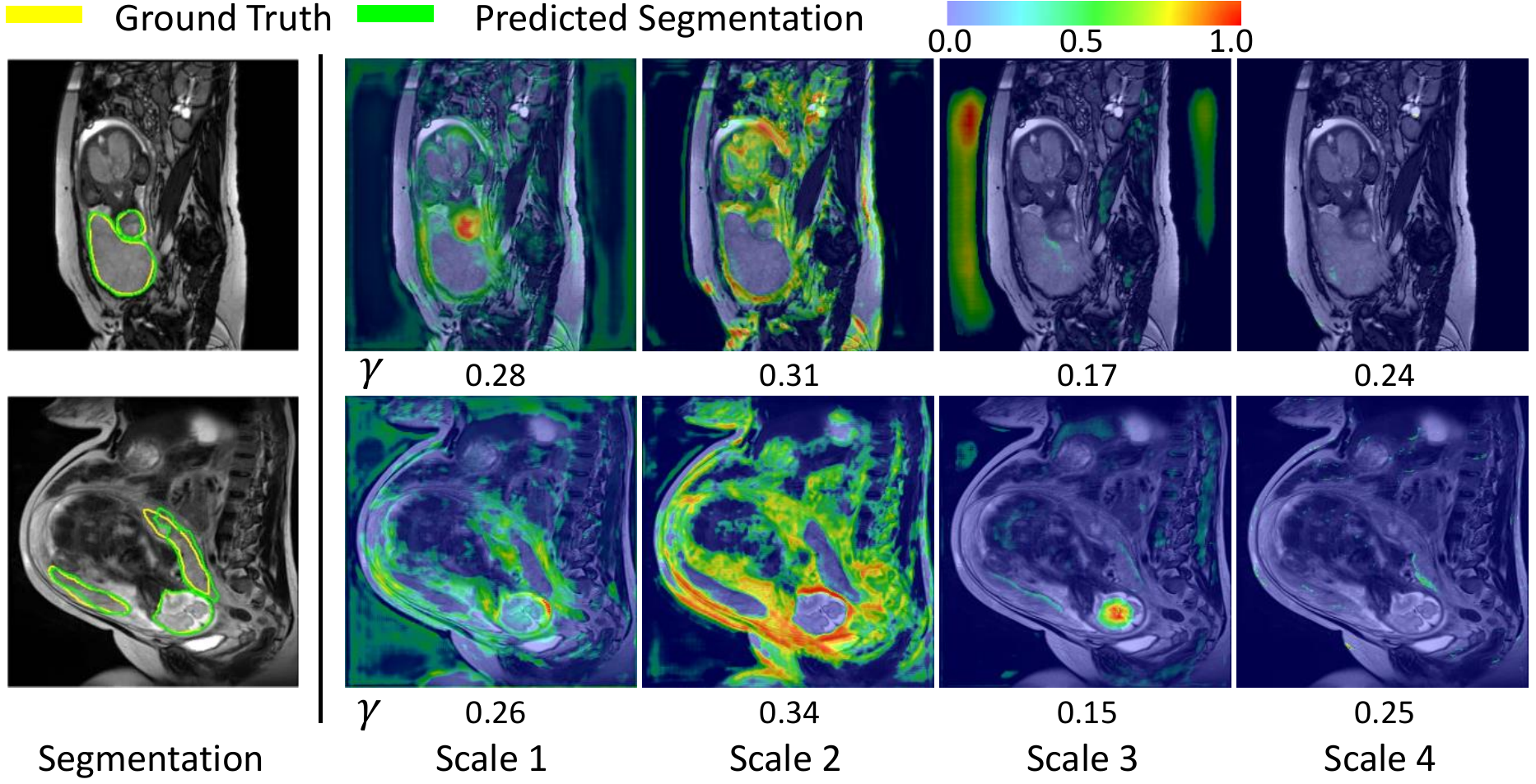}
    \caption{Visualization of scale attention weight maps on fetal MRI. Warmer color represents higher attention coefficient values. $\gamma$ means the global channel-wise scale attention coefficient.}
    \label{fig6_3:scale_atten_weight_mri}
\end{figure}
\subsubsection{Comparison of Partial and Comprehensive Attention}
Similar to Section~\ref{partial attention skin}, we compared comprehensive attentions with partial attentions in the task of segmenting fetal brain and placenta from fetal MRI. From Table~\ref{tab5:full_partial_table_mri}, we find that models combining two of the three attention mechanism basically outperform variants that using a single attention mechanism. SA + CA gets the highest scores among three binary-attention methods, which can achieve Dice scores of 86.68\% for the placenta and 95.42\% for the fetal brain. Furthermore, our proposed CA-Net outperforms all these binary-attention methods, achieving Dice scores of 87.08\% for the placenta and 95.88\% for the fetal brain, respectively. The ASSD value of CA-Net is lower than those of other methods.
\begin{table}[htb]
    \small
    \vspace{-0.4cm}
    \setlength{\abovecaptionskip}{0.2cm}
    \setlength{\belowcaptionskip}{-0.4cm}
    \centering
    \caption{Quantitative comparison of partial and comprehensive attention methods for fetal MRI segmentation. SA, CA and LA are our proposed spatial, channel and scale attention modules respectively.}
    \label{tab5:full_partial_table_mri}
    \scalebox{0.84}{\begin{tabular}{l|c|c|c|c}
        \hline
        \multicolumn{1}{l}{Network} & \multicolumn{2}{|c}{Placenta} & \multicolumn{2}{|c}{Fetal Brain}\\
        \cline{2-5}
        \multicolumn{1}{l|}{} & Dice(\%) & ASSD(pix) & Dice(\%) & ASSD(pix) \\ \hline
        Baseline & 84.79$\pm$8.45 & 0.77$\pm$0.95 & 93.20$\pm$5.96 & 0.38$\pm$0.92 \\
        SA & 85.65$\pm$6.19 & 0.58$\pm$0.43 & 95.47$\pm$2.43 & 0.30$\pm$0.49 \\
        LA & 86.21$\pm$5.96 & 0.52$\pm$0.58 & 95.18$\pm$3.22 & 0.27$\pm$0.59 \\
        CA & 86.65$\pm$5.99 & 0.52$\pm$0.40 & 95.39$\pm$2.66 & 0.28$\pm$0.39 \\
        SA+LA & 86.20$\pm$6.26 & 0.54$\pm$0.42 & 94.59$\pm$3.14 & 0.35$\pm$0.53 \\
        CA+LA & 86.50$\pm$6.89 & 0.47$\pm$0.29 & 95.29$\pm$3.10 & 0.32$\pm$0.59 \\
        SA+CA & 86.68$\pm$4.82 & 0.48$\pm$0.42 & 95.42$\pm$2.44 & 0.25$\pm$0.45 \\
        \textbf{CA-Net(Ours)} &\textbf{87.08$\pm$6.07} & \textbf{0.52$\pm$0.39} & \textbf{95.88$\pm$2.07} & \textbf{0.16$\pm$0.26} \\ \hline
    \end{tabular}}
\end{table}
\begin{table*}[htb]
    \small
    \vspace{-0.4cm}
    \setlength{\abovecaptionskip}{0.2cm}
    \setlength{\belowcaptionskip}{-0.6cm}
    \centering
    \caption{Quantitative evaluations of the state-of-the-art methods and our proposed FA-Net for fetal MRI segmentation on three views (axial, coronal and sagittal). Inf-T means the inference time on whole test dataset. E-able means the method is explainable.}
    \label{tab4:mri_fetal_table_mri}
    \scalebox{0.8}{\begin{tabular}{l|c|c|c|c|c|c|c|c|c|c}
        \hline
         & & & \multicolumn{8}{c}{Placenta} \\
        \cline{4-11}
        \multicolumn{1}{l|}{Network}& \multicolumn{1}{c|}{Para/Inf-T} & \multicolumn{1}{c|}{E-able} & \multicolumn{2}{c|}{Axial} & \multicolumn{2}{c|}{Coronal} & \multicolumn{2}{c|}{Sagittal} & \multicolumn{2}{c}{Whole} \\
        \cline{4-11}
        & & & Dice(\%) & ASSD(pix) & Dice(\%) & ASSD(pix) & Dice(\%) & ASSD(pix) & Dice(\%) & ASSD(pix) \\ \hline
        Baseline(U-Net~\cite{ronneberger2015u}) & 1.9M/0.9ms &$\times$ & 85.37$\pm$7.12 & 0.38$\pm$0.18 & 83.32$\pm$5.41 & 0.90$\pm$0.88 & 85.28$\pm$10.48 & 1.06$\pm$1.33 &
        84.79$\pm$8.45 & 0.77$\pm$0.95 \\
        Attention U-Net~\cite{Oktay2018a} & 2.1M/1.1ms &$\surd$& 86.00$\pm$5.94 & \textbf{0.44$\pm$0.29} & 84.56$\pm$7.47 & 0.71$\pm$0.45 & 85.74$\pm$10.20 & 1.00$\pm$1.27 & 85.52$\pm$7.89 & 0.72$\pm$0.83\\
        DenseASPP~\cite{yang2018denseaspp} & 8.3M/3.1ms &$\times$& 84.29$\pm$5.84 & 0.67$\pm$0.70 & 82.11$\pm$7.28 & 0.60$\pm$0.21 & 84.92$\pm$10.02 & 1.06$\pm$1.17 & 83.93$\pm$7.84 & 0.80$\pm$0.84 \\
        DeepLabv3+(DRN) & 40.7M/1.6ms &$\times$& 85.79$\pm$6.00 & 0.60$\pm$0.56 & 82.34$\pm$6.69 & 0.73$\pm$0.46 & 85.98$\pm$7.11 & 0.90$\pm$0.58 & 84.91$\pm$6.59 & 0.75$\pm$0.54 \\
        RefineNet~\cite{lin2017refinenet} & 46.3M/2.2ms &$\times$& 86.30$\pm$5.72 & 0.55$\pm$0.48 & 83.25$\pm$5.55 & 0.64$\pm$0.40 & 86.67$\pm$7.04 & 0.88$\pm$0.79 & 85.60$\pm$6.18 & 0.70$\pm$0.60 \\
        DeepLabv3+~\cite{chen2018encoder} & 54.7M/3.4ms &$\times$& 86.34$\pm$6.30 & 0.46$\pm$0.50 & 83.38$\pm$7.23 & 0.54$\pm$0.37 & 86.90$\pm$7.84 & 0.58$\pm$0.49 & 85.76$\pm$7.17 & 0.57$\pm$0.45 \\
        \textbf{CA-Net(Ours)} & 2.8M/1.5ms &$\surd$& \textbf{87.72$\pm$4.43} & 0.53$\pm$0.42 & \textbf{85.10$\pm$5.83} & \textbf{0.52$\pm$0.35} & \textbf{87.88$\pm$7.67} & \textbf{0.50$\pm$0.45} & \textbf{87.08$\pm$6.07} & \textbf{0.52$\pm$0.39} \\ \hline
        & & & \multicolumn{8}{c}{Brain} \\
        \cline{4-11}
        \multicolumn{1}{l|}{Network}& \multicolumn{1}{c|}{Para/Inf-T} & \multicolumn{1}{c|}{E-able} & \multicolumn{2}{c|}{Axial} & \multicolumn{2}{c|}{Coronal} & \multicolumn{2}{c|}{Sagittal} & \multicolumn{2}{c}{Whole} \\
        \cline{4-11}
        & & & Dice(\%) & ASSD(pix) & Dice(\%) & ASSD(pix) & Dice(\%) & ASSD(pix) & Dice(\%) & ASSD(pix) \\ \hline
        Baseline(U-Net~\cite{ronneberger2015u}) & 1.9M/0.9ms & $\times$ & 91.76$\pm$7.44 & 0.22$\pm$0.25 & 95.67$\pm$2.28 & 0.15$\pm$0.15 & 92.84$\pm$6.04 & 0.71$\pm$1.48 & 93.20$\pm$5.96 & 0.38$\pm$0.92 \\
        Attention U-Net~\cite{Oktay2018a} & 2.1M/1.1ms &$\surd$& 93.27$\pm$3.21 & 0.20$\pm$0.16 & 94.79$\pm$5.49 & 0.57$\pm$1.36 & 94.85$\pm$3.73 & 0.19$\pm$0.21 & 94.25$\pm$4.03 & 0.30$\pm$0.70\\
        DenseASPP~\cite{yang2018denseaspp} & 8.3M/3.1ms &$\times$& 91.08$\pm$5.52 & 0.24$\pm$0.21 & 93.42$\pm$4.82 & 0.25$\pm$0.23 & 92.02$\pm$6.04 & 0.68$\pm$1.76 & 92.08$\pm$5.43 & 0.41$\pm$1.09 \\
        DeepLabv3+(DRN) & 40.7M/1.6ms &$\times$& 92.63$\pm$3.00 & 0.33$\pm$0.42 & 94.94$\pm$2.49 & 0.14$\pm$0.09 & 92.36$\pm$0.09 & 0.94$\pm$2.75 & 93.10$\pm$5.90 & 0.51$\pm$1.70 \\
        RefineNet~\cite{lin2017refinenet} & 46.3M/2.2ms &$\times$& 94.04$\pm$1.85 & 0.25$\pm$0.45 & 91.90$\pm$11.10 & 0.25$\pm$0.41 & 90.36$\pm$10.91 & 0.81$\pm$2.09 & 92.05$\pm$8.77 & 0.46$\pm$1.32 \\
        DeepLabv3+~\cite{chen2018encoder} & 54.7M/3.4ms &$\times$& 93.70$\pm$4.17 & \textbf{0.19$\pm$0.31} & 95.50$\pm$2.60 & \textbf{0.09$\pm$0.03} & 94.97$\pm$4.27 & 0.58$\pm$1.24 & 94.65$\pm$3.87 & 0.31$\pm$0.78 \\
        \textbf{CA-Net(Ours)} & 2.8M/1.5ms &$\surd$& \textbf{95.51$\pm$2.66} & 0.21$\pm$0.23 & \textbf{96.52$\pm$1.61} & 0.09$\pm$0.05 & \textbf{95.68$\pm$3.49} & \textbf{0.09$\pm$0.09} & \textbf{95.88$\pm$2.07} & \textbf{0.16$\pm$0.26}\\ \hline
    \end{tabular}}
\end{table*}
\subsubsection{Comparison of State-of-the-Art Frameworks}
\begin{figure}[htb]
    \centering
    \vspace{-0.2cm}
    \setlength{\abovecaptionskip}{-0cm}
    \setlength{\belowcaptionskip}{-0.2cm}
    \includegraphics[width=0.5\textwidth]{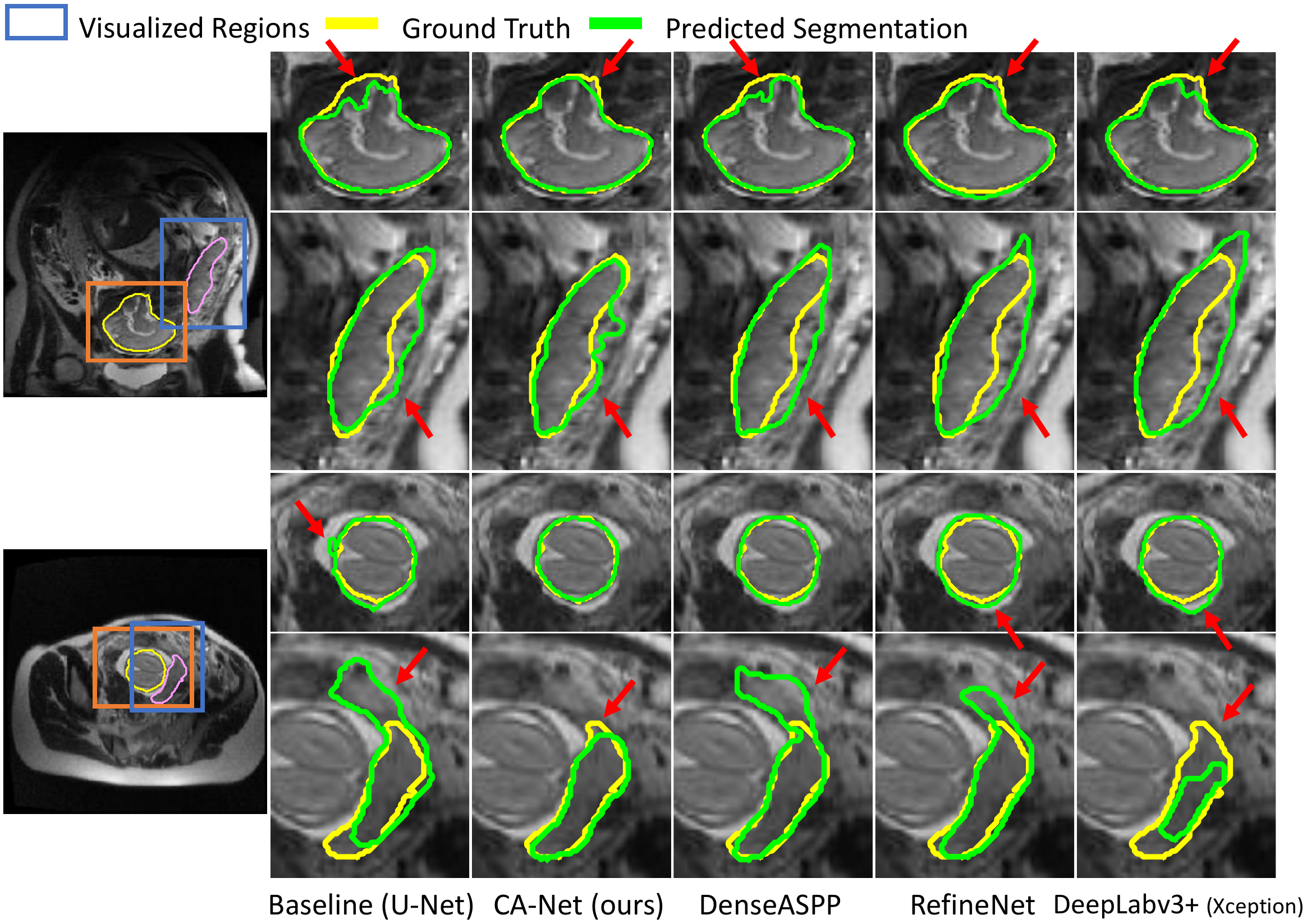}
    \caption{Visual comparison of our proposed CA-Net with the state-of-the-art segmentation methods for fetal brain and placenta segmentation from MRI. Red arrows highlight the mis-segmented regions.}
    \label{fig8:state_of_the_art_result_mri}
\end{figure}

We also compared our CA-Net with the state-of-the-art methods and their variants as implemented in section~\ref{partial attention skin}. The segmentation performance on images in axial, sagittal and coronal views was measured respectively. A quantitative evaluation of these methods for fetal MRI segmentation is listed in Table~\ref{tab4:mri_fetal_table_mri}. We observe that our proposed CA-Net obtained better Dice scores than the others in all the three views. Our CA-Net can improve the Dice scores by $2.35\%$, $1.78\%$, and $2.60\%$ for placenta segmentation and $3.75\%$, $0.85\%$, and $2.84\%$ for fetal brain segmentation in three views compared with U-Net, respectively, surpassing the existing attention method and the state-of-the-art segmentation methods. In addition, for the average Dice and ASSD values across the three views, CA-Net outperformed the others.
Meanwhile, CA-Net has a much smaller model size compared with RefineNet~\cite{lin2017refinenet} and Deeplabv3+~\cite{chen2018encoder}, which leads to lower computational cost for training and inference. For fetal MRI segmentation, the average inference time per image for our CA-Net was 1.5ms, compared with 3.4ms and 2.2ms by DeepLabv3+ and RefineNet, respectively. 
Qualitative results in Fig.~\ref{fig8:state_of_the_art_result_mri} also show that CA-Net performs noticeably better than the baseline and the other methods for fetal MRI segmentation. In dealing with the complex shapes as shown in the first and fifth rows of Fig.~\ref{fig8:state_of_the_art_result_mri}, as well as the blurry boundary in the second row, CA-Net performs more closely to the authentic boundary than the other methods. Note that visualization of the spatial and scale attentions as show in Fig.~\ref{fig5:spatial_atten_result_with_atten_mri} and Fig.~\ref{fig6_3:scale_atten_weight_mri} helps to interpret the decision of our CA-Net, but such explainability is not provided by DeepLabv3+, RefineNet and DenseASPP.

\section{Discussion and Conclusion}
\label{}
For a medical image segmentation task, some targets such as lesions may have a large variation of position, shape and scale, enabling the networks to be aware of the object’s spatial position and size is important for accurate segmentation. In addition, convolution neural networks generate feature maps with a large number of channels, and concatenation of feature maps with different semantic information or from different scales are often used. Paying an attention to the most relevant channels and scales is an effective way to improve the segmentation performance. Using scale attention to adaptively make use of the features in different scales would have advantages in dealing with objects with a variation of scales. To take these advantages simultaneously, we take a comprehensive use of these complementary attention mechanisms, and our results show that CA-Net helps to obtain more accurate segmentation with only few parameters.

For explainable CNN, previous works like CAM~\cite{zhou2016learning} and GBP~\cite{springenberg2014striving} mainly focused on image classification tasks and they only consider the spatial information for explaining the CNN's prediction. In addition, they are post-hoc methods that require additional computations after a forward pass prediction to interpret the prediction results. Differently from these methods, CA-Net gives a comprehensive interpretation of how each spatial position, feature map channel and scale is used for the prediction in segmentation tasks. What's more, we obtain these attention coefficients in a single forward pass and require no additional computations. By visualizing the attention maps in different aspects as show in Fig.~\ref{fig5:spatial_atten_result_with_atten_skin} and Fig.~\ref{fig6_2:scale_atten_weight}, we can better understand how the network works, which has a potential to helps us improve the design of CNNs.

We have done experiments on two different image domains, i.e., RGB image and fetal MRI. These two are representative image domains, and in both cases our CA-Net has a considerable segmentation improvement compared with U-Net, which shows that the CA-Net has competing performance for different segmentation tasks in different modalities. It is of interest to apply our CA-Net to other image modalities such as the Ultrasound and other anatomies in the future.

In this work, we have investigated three main types of attentions associated with segmentation targets in various positions and scales. Recently, some other types of attentions have also been proposed in the literature, such as attention to parallel convolution kernels~\cite{chen2020dynamic}. However, using multiple parallel convolution kernels will increase the model complexity.

Most of the attention blocks in our CA-Net are in the decoder. This is mainly because that the encoder acts as a feature extractor that is exploited to obtain enough candidate features. Applying attention at the encoder may lead some potentially useful features to be suppressed at an early stage. Therefore, we use the attention blocks in the decoder to highlight relevant features from all the candidate features. Specifically, following~\cite{Oktay2018a}, the spatial attention is designed to use high-level semantic features in the decoder to calibrate low-level features in the encoder, so they are used at the skip connections after the encoder. The scale attention is designed to better fuse the raw semantic predictions that are obtained in the decoder, which should naturally be placed at the end of the network. For channel attentions, we tried to place them at different positions of the network, and found that placing them in the decoder is better than in the encoder. As shown in Table II, all the channel attention variants outperformed the baseline U-Net. However, using channel attention only in the decoder outperformed the variants with channel attention in the encoder. The reason may be that the encoding phase needs to maintain enough feature information, which confirms our assumption that suppressing some features at an early stage will limit the model’s performance. However, some other attentions~\cite{chen2020dynamic} might be useful in the encoder, which will be investigated in the future.


Differently from previous works that mainly focus on improving the segmentation accuracy while hard to explain, we aim to design a network with good comprehensive property, including high segmentation accuracy, efficiency and explainability at the same time. Indeed, the segmentation accuracy of our CA-Net is competing: It leads to a significant improvement of Dice compared with the U-Net ($92.08\%$ VS $87.77\%$) for skin lesion. Compared with state-of-the-art DeepLabv3+ and RefineNet, our CA-Net achieved very close segmentation accuracy with around 15 times fewer parameters. What’s more, CA-Net is easy to explain as shown in Fig.~\ref{fig5:spatial_atten_result_with_atten_skin},~\ref{fig6_2:scale_atten_weight},~\ref{fig5:spatial_atten_result_with_atten_mri}, and~\ref{fig6_3:scale_atten_weight_mri}, but DeepLabv3+ and RefineNet have poor explainability on how they localize the target region, recognize the scale and determine the useful features. Meanwhile, in fetal MRI segmentation, experimental results from Table X shows that our CA-Net has a considerable improvement compared with U-Net (Dice was $87.08\%$ VS $84.79\%$), and it outperforms the state-of-the-art methods in all the three views. Therefore, the superiority of our CA-Net is that it could achieve high explainability and efficiency than state-of-the-art methods while maintaining comparable or even better accuracy.

In the skin lesion segmentation task, we observe that our CA-Net leads to slightly inferior performance than Deeplabv3+, which is however without significant difference. We believe the reason is that Deeplabv3+ is mainly designed for natural image segmentation task, and the dermoscopic skin images are color images, which has a similar distribution of intensity to natural images. 
However, compared to Deeplabv3+, our CA-Net can achieve comparable performance, and it has higher explainability and 15 times fewer parameters, leading to higher computational efficiency. In the fetal MRI segmentation task, our CA-Net has distinctly higher accuracy than those state-of-the-art methods, which shows the effectiveness and good explainability of our method.


In conclusion, we propose a comprehensive attention-based convolutional neural network (CA-Net) that learns to make extensive use of multiple attentions for better performance and explainability of medical image segmentation. We enable the network to adaptively pay attention to spatial positions, feature channels and object scales at the same time.
Motivated by existing spatial and channel attention methods, we make further improvements to enhance the network's ability to focus on areas of interest. We propose a novel scale attention module implicitly emphasizing the most salient scales to obtain multiple-scale features.
Experimental results show that compared with the state-of-the-art semantic segmentation models like Deeplabv3+, our CA-Net obtains comparable and even higher accuracy for medical image segmentation with a much smaller model size. Most importantly, CA-Net gains a good model explainability which is important for understanding how the network works, and has a potential to improve clinicians’ acceptance and trust on predictions given by an artificial intelligence algorithm. Our proposed multiple attention modules can be easily plugged into most semantic segmentation networks.
In the future, the method can be easily extended to segment 3D images.


%









\bibliographystyle{IEEEtran}
\end{document}